\documentclass[lettersize,journal]{IEEEtran}

\usepackage{cite}
\usepackage[T1]{fontenc}
\usepackage{graphicx}
\usepackage{amssymb}
\usepackage{amsmath}
\usepackage{amsthm}
\usepackage{subfigure}
\usepackage{booktabs} 
\usepackage{multirow}
\usepackage{microtype}
\usepackage{balance}
\usepackage{xcolor}
\usepackage{xfrac}    
\usepackage{algorithm}
\usepackage{setspace}
\usepackage{algorithmicx}
\usepackage{algpseudocode}
\algrenewcommand\algorithmicindent{0.7em}
\usepackage{xpatch}
\usepackage{hyperref}

\usepackage{tikz}

\usepackage{amssymb}
\usepackage{amsfonts}
\usepackage{mathrsfs}
\usepackage{xspace}
\usepackage{bm}
\usepackage{upgreek}

\newcommand{\safemath}[2]{\newcommand{#1}{\ensuremath{#2}\xspace}}

\safemath{\bma}{\mathbf{a}}
\safemath{\bmb}{\mathbf{b}}
\safemath{\bmc}{\mathbf{c}}
\safemath{\bmd}{\mathbf{d}}
\safemath{\bme}{\mathbf{e}}
\safemath{\bmf}{\mathbf{f}}
\safemath{\bmg}{\mathbf{g}}
\safemath{\bmh}{\mathbf{h}}
\safemath{\bmi}{\mathbf{i}}
\safemath{\bmj}{\mathbf{j}}
\safemath{\bmk}{\mathbf{k}}
\safemath{\bml}{\mathbf{l}}
\safemath{\bmm}{\mathbf{m}}
\safemath{\bmn}{\mathbf{n}}
\safemath{\bmo}{\mathbf{o}}
\safemath{\bmp}{\mathbf{p}}
\safemath{\bmq}{\mathbf{q}}
\safemath{\bmr}{\mathbf{r}}
\safemath{\bms}{\mathbf{s}}
\safemath{\bmt}{\mathbf{t}}
\safemath{\bmu}{\mathbf{u}}
\safemath{\bmv}{\mathbf{v}}
\safemath{\bmw}{\mathbf{w}}
\safemath{\bmx}{\mathbf{x}}
\safemath{\bmy}{\mathbf{y}}
\safemath{\bmz}{\mathbf{z}}
\safemath{\bmzero}{\mathbf{0}}
\safemath{\bmone}{\mathbf{1}}

\bmdefine{\biad}{a}
\bmdefine{\bibd}{b}
\bmdefine{\bicd}{c}
\bmdefine{\bidd}{d}
\bmdefine{\bied}{e}
\bmdefine{\bifd}{f}
\bmdefine{\bigd}{g}
\bmdefine{\bihd}{h}
\bmdefine{\biid}{i}
\bmdefine{\bijd}{j}
\bmdefine{\bikd}{k}
\bmdefine{\bild}{l}
\bmdefine{\bimd}{m}
\bmdefine{\bind}{n}
\bmdefine{\biod}{o}
\bmdefine{\bipd}{p}
\bmdefine{\biqd}{q}
\bmdefine{\bird}{r}
\bmdefine{\bisd}{s}
\bmdefine{\bitd}{t}
\bmdefine{\biud}{u}
\bmdefine{\bivd}{v}
\bmdefine{\biwd}{w}
\bmdefine{\bixd}{x}
\bmdefine{\biyd}{y}
\bmdefine{\bizd}{z}

\bmdefine{\bixid}{\xi}
\bmdefine{\bilambdad}{\lambda}
\bmdefine{\bimud}{\mu}
\bmdefine{\bithetad}{\theta}
\bmdefine{\biphid}{\phi}
\bmdefine{\bideltad}{\delta}

\safemath{\bmia}{\biad}
\safemath{\bmib}{\bibd}
\safemath{\bmic}{\bicd}
\safemath{\bmid}{\bidd}
\safemath{\bmie}{\bied}
\safemath{\bmif}{\bifd}
\safemath{\bmig}{\bigd}
\safemath{\bmih}{\bihd}
\safemath{\bmii}{\biid}
\safemath{\bmij}{\bijd}
\safemath{\bmik}{\bikd}
\safemath{\bmil}{\bild}
\safemath{\bmim}{\bimd}
\safemath{\bmin}{\bind}
\safemath{\bmio}{\biod}
\safemath{\bmip}{\bipd}
\safemath{\bmiq}{\biqd}
\safemath{\bmir}{\bird}
\safemath{\bmis}{\bisd}
\safemath{\bmit}{\bitd}
\safemath{\bmiu}{\biud}
\safemath{\bmiv}{\bivd}
\safemath{\bmiw}{\biwd}
\safemath{\bmix}{\bixd}
\safemath{\bmiy}{\biyd}
\safemath{\bmiz}{\bizd}

\safemath{\bmxi}{\bixid}
\safemath{\bmlambda}{\bilambdad}
\safemath{\bmmu}{\bimud}
\safemath{\bmtheta}{\bithetad}
\safemath{\bmphi}{\biphid}
\safemath{\bmdelta}{\bideltad}

\safemath{\bA}{\mathbf{A}}
\safemath{\bB}{\mathbf{B}}
\safemath{\bC}{\mathbf{C}}
\safemath{\bD}{\mathbf{D}}
\safemath{\bE}{\mathbf{E}}
\safemath{\bF}{\mathbf{F}}
\safemath{\bG}{\mathbf{G}}
\safemath{\bH}{\mathbf{H}}
\safemath{\bI}{\mathbf{I}}
\safemath{\bJ}{\mathbf{J}}
\safemath{\bK}{\mathbf{K}}
\safemath{\bL}{\mathbf{L}}
\safemath{\bM}{\mathbf{M}}
\safemath{\bN}{\mathbf{N}}
\safemath{\bO}{\mathbf{O}}
\safemath{\bP}{\mathbf{P}}
\safemath{\bQ}{\mathbf{Q}}
\safemath{\bR}{\mathbf{R}}
\safemath{\bS}{\mathbf{S}}
\safemath{\bT}{\mathbf{T}}
\safemath{\bU}{\mathbf{U}}
\safemath{\bV}{\mathbf{V}}
\safemath{\bW}{\mathbf{W}}
\safemath{\bX}{\mathbf{X}}
\safemath{\bY}{\mathbf{Y}}
\safemath{\bZ}{\mathbf{Z}}

\safemath{\bZero}{\mathbf{0}}
\safemath{\bOne}{\mathbf{1}}
\safemath{\bDelta}{\mathbf{\Delta}}
\safemath{\bLambda}{\mathbf{\Lambda}}
\safemath{\bPhi}{\mathbf{\Upphi}}
\safemath{\bSigma}{\mathbf{\Upsigma}}
\safemath{\bOmega}{\mathbf{\Upomega}}
\safemath{\bTheta}{\mathbf{\Uptheta}}

\bmdefine{\biAd}{A}
\bmdefine{\biBd}{B}
\bmdefine{\biCd}{C}
\bmdefine{\biDd}{D}
\bmdefine{\biEd}{E}
\bmdefine{\biFd}{F}
\bmdefine{\biGd}{G}
\bmdefine{\biHd}{H}
\bmdefine{\biId}{I}
\bmdefine{\biJd}{J}
\bmdefine{\biKd}{K}
\bmdefine{\biLd}{L}
\bmdefine{\biMd}{M}
\bmdefine{\biOd}{N}
\bmdefine{\biPd}{O}
\bmdefine{\biQd}{P}
\bmdefine{\biRd}{R}
\bmdefine{\biSd}{S}
\bmdefine{\biTd}{T}
\bmdefine{\biUd}{U}
\bmdefine{\biVd}{V}
\bmdefine{\biWd}{W}
\bmdefine{\biXd}{X}
\bmdefine{\biYd}{Y}
\bmdefine{\biZd}{Z}

\bmdefine{\biDelta}{\Delta}
\bmdefine{\biLambda}{\Lambda}
\bmdefine{\biPhi}{\Phi}
\bmdefine{\biSigma}{\Sigma}
\bmdefine{\biOmega}{\Omega}
\bmdefine{\biTheta}{\Theta}

\safemath{\bimA}{\biAd}
\safemath{\bimB}{\biBd}
\safemath{\bimC}{\biCd}
\safemath{\bimD}{\biDd}
\safemath{\bimE}{\biEd}
\safemath{\bimF}{\biFd}
\safemath{\bimG}{\biGd}
\safemath{\bimH}{\biHd}
\safemath{\bimI}{\biId}
\safemath{\bimJ}{\biJd}
\safemath{\bimK}{\biKd}
\safemath{\bimL}{\biLd}
\safemath{\bimM}{\biMd}
\safemath{\bimN}{\biNd}
\safemath{\bimO}{\biOd}
\safemath{\bimP}{\biPd}
\safemath{\bimQ}{\biQd}
\safemath{\bimR}{\biRd}
\safemath{\bimS}{\biSd}
\safemath{\bimT}{\biTd}
\safemath{\bimU}{\biUd}
\safemath{\bimV}{\biVd}
\safemath{\bimW}{\biWd}
\safemath{\bimX}{\biXd}
\safemath{\bimY}{\biYd}
\safemath{\bimZ}{\biZd}

\safemath{\bimDelta}{\biDelta}
\safemath{\bimLambda}{\biLambda}
\safemath{\bimPhi}{\biPhi}
\safemath{\bimSigma}{\biSigma}
\safemath{\bimOmega}{\biOmega}
\safemath{\bimTheta}{\biTheta}

\safemath{\setA}{\mathcal{A}}
\safemath{\setB}{\mathcal{B}}
\safemath{\setC}{\mathcal{C}}
\safemath{\setD}{\mathcal{D}}
\safemath{\setE}{\mathcal{E}}
\safemath{\setF}{\mathcal{F}}
\safemath{\setG}{\mathcal{G}}
\safemath{\setH}{\mathcal{H}}
\safemath{\setI}{\mathcal{I}}
\safemath{\setJ}{\mathcal{J}}
\safemath{\setK}{\mathcal{K}}
\safemath{\setL}{\mathcal{L}}
\safemath{\setM}{\mathcal{M}}
\safemath{\setN}{\mathcal{N}}
\safemath{\setO}{\mathcal{O}}
\safemath{\setP}{\mathcal{P}}
\safemath{\setQ}{\mathcal{Q}}
\safemath{\setR}{\mathcal{R}}
\safemath{\setS}{\mathcal{S}}
\safemath{\setT}{\mathcal{T}}
\safemath{\setU}{\mathcal{U}}
\safemath{\setV}{\mathcal{V}}
\safemath{\setW}{\mathcal{W}}
\safemath{\setX}{\mathcal{X}}
\safemath{\setY}{\mathcal{Y}}
\safemath{\setZ}{\mathcal{Z}}
\safemath{\emptySet}{\varnothing}

\safemath{\colA}{\mathscr{A}}
\safemath{\colB}{\mathscr{B}}
\safemath{\colC}{\mathscr{C}}
\safemath{\colD}{\mathscr{D}}
\safemath{\colE}{\mathscr{E}}
\safemath{\colF}{\mathscr{F}}
\safemath{\colG}{\mathscr{G}}
\safemath{\colH}{\mathscr{H}}
\safemath{\colI}{\mathscr{I}}
\safemath{\colJ}{\mathscr{J}}
\safemath{\colK}{\mathscr{K}}
\safemath{\colL}{\mathscr{L}}
\safemath{\colM}{\mathscr{M}}
\safemath{\colN}{\mathscr{N}}
\safemath{\colO}{\mathscr{O}}
\safemath{\colP}{\mathscr{P}}
\safemath{\colQ}{\mathscr{Q}}
\safemath{\colR}{\mathscr{R}}
\safemath{\colS}{\mathscr{S}}
\safemath{\colT}{\mathscr{T}}
\safemath{\colU}{\mathscr{U}}
\safemath{\colV}{\mathscr{V}}
\safemath{\colW}{\mathscr{W}}
\safemath{\colX}{\mathscr{X}}
\safemath{\colY}{\mathscr{Y}}
\safemath{\colZ}{\mathscr{Z}}

\safemath{\opA}{\mathbb{A}}
\safemath{\opB}{\mathbb{B}}
\safemath{\opC}{\mathbb{C}}
\safemath{\opD}{\mathbb{D}}
\safemath{\opE}{\mathbb{E}}
\safemath{\opF}{\mathbb{F}}
\safemath{\opG}{\mathbb{G}}
\safemath{\opH}{\mathbb{H}}
\safemath{\opI}{\mathbb{I}}
\safemath{\opJ}{\mathbb{J}}
\safemath{\opK}{\mathbb{K}}
\safemath{\opL}{\mathbb{L}}
\safemath{\opM}{\mathbb{M}}
\safemath{\opN}{\mathbb{N}}
\safemath{\opO}{\mathbb{O}}
\safemath{\opP}{\mathbb{P}}
\safemath{\opQ}{\mathbb{Q}}
\safemath{\opR}{\mathbb{R}}
\safemath{\opS}{\mathbb{S}}
\safemath{\opT}{\mathbb{T}}
\safemath{\opU}{\mathbb{U}}
\safemath{\opV}{\mathbb{V}}
\safemath{\opW}{\mathbb{W}}
\safemath{\opX}{\mathbb{X}}
\safemath{\opY}{\mathbb{Y}}
\safemath{\opZ}{\mathbb{Z}}
\safemath{\opZero}{\mathbb{O}}
\safemath{\identityop}{\opI}

\safemath{\veca}{\bma}
\safemath{\vecb}{\bmb}
\safemath{\vecc}{\bmc}
\safemath{\vecd}{\bmd}
\safemath{\vece}{\bme}
\safemath{\vecf}{\bmf}
\safemath{\vecg}{\bmg}
\safemath{\vech}{\bmh}
\safemath{\veci}{\bmi}
\safemath{\vecj}{\bmj}
\safemath{\veck}{\bmk}
\safemath{\vecl}{\bml}
\safemath{\vecm}{\bmm}
\safemath{\vecn}{\bmn}
\safemath{\veco}{\bmo}
\safemath{\vecp}{\bmp}
\safemath{\vecq}{\bmq}
\safemath{\vecr}{\bmr}
\safemath{\vecs}{\bms}
\safemath{\vect}{\bmt}
\safemath{\vecu}{\bmu}
\safemath{\vecv}{\bmv}
\safemath{\vecw}{\bmw}
\safemath{\vecx}{\bmx}
\safemath{\vecy}{\bmy}
\safemath{\vecz}{\bmz}

\safemath{\veczero}{\bmzero}
\safemath{\vecone}{\bmone}
\safemath{\vecxi}{\bmxi}
\safemath{\veclambda}{\bmlambda}
\safemath{\vecmu}{\bmmu}
\safemath{\vectheta}{\bmtheta}
\safemath{\vecphi}{\bmphi}
\safemath{\vecdelta}{\bmdelta}

\safemath{\matA}{\bA}
\safemath{\matB}{\bB}
\safemath{\matC}{\bC}
\safemath{\matD}{\bD}
\safemath{\matE}{\bE}
\safemath{\matF}{\bF}
\safemath{\matG}{\bG}
\safemath{\matH}{\bH}
\safemath{\matI}{\bI}
\safemath{\matJ}{\bJ}
\safemath{\matK}{\bK}
\safemath{\matL}{\bL}
\safemath{\matM}{\bM}
\safemath{\matN}{\bN}
\safemath{\matO}{\bO}
\safemath{\matP}{\bP}
\safemath{\matQ}{\bQ}
\safemath{\matR}{\bR}
\safemath{\matS}{\bS}
\safemath{\matT}{\bT}
\safemath{\matU}{\bU}
\safemath{\matV}{\bV}
\safemath{\matW}{\bW}
\safemath{\matX}{\bX}
\safemath{\matY}{\bY}
\safemath{\matZ}{\bZ}
\safemath{\matzero}{\bmzero}

\safemath{\matDelta}{\bDelta}
\safemath{\matLambda}{\bLambda}
\safemath{\matPhi}{\bPhi}
\safemath{\matSigma}{\bSigma}
\safemath{\matOmega}{\bOmega}
\safemath{\matTheta}{\bTheta}

\safemath{\matidentity}{\matI}
\safemath{\matone}{\matO}

\safemath{\rnda}{A}
\safemath{\rndb}{B}
\safemath{\rndc}{C}
\safemath{\rndd}{D}
\safemath{\rnde}{E}
\safemath{\rndf}{F}
\safemath{\rndg}{G}
\safemath{\rndh}{H}
\safemath{\rndi}{I}
\safemath{\rndj}{J}
\safemath{\rndk}{K}
\safemath{\rndl}{L}
\safemath{\rndm}{M}
\safemath{\rndn}{N}
\safemath{\rndo}{O}
\safemath{\rndp}{P}
\safemath{\rndq}{Q}
\safemath{\rndr}{R}
\safemath{\rnds}{S}
\safemath{\rndt}{T}
\safemath{\rndu}{U}
\safemath{\rndv}{V}
\safemath{\rndw}{W}
\safemath{\rndx}{X}
\safemath{\rndy}{Y}
\safemath{\rndz}{Z}

\safemath{\rveca}{\bimA}
\safemath{\rvecb}{\bimB}
\safemath{\rvecc}{\bimC}
\safemath{\rvecd}{\bimD}
\safemath{\rvece}{\bimE}
\safemath{\rvecf}{\bimF}
\safemath{\rvecg}{\bimG}
\safemath{\rvech}{\bimH}
\safemath{\rveci}{\bimI}
\safemath{\rvecj}{\bimJ}
\safemath{\rveck}{\bimK}
\safemath{\rvecl}{\bimL}
\safemath{\rvecm}{\bimM}
\safemath{\rvecn}{\bimN}
\safemath{\rveco}{\bomO}
\safemath{\rvecp}{\bimP}
\safemath{\rvecq}{\bimQ}
\safemath{\rvecr}{\bimR}
\safemath{\rvecs}{\bimS}
\safemath{\rvect}{\bimT}
\safemath{\rvecu}{\bimU}
\safemath{\rvecv}{\bimV}
\safemath{\rvecw}{\bimW}
\safemath{\rvecx}{\bimX}
\safemath{\rvecy}{\bimY}
\safemath{\rvecz}{\bimZ}

\safemath{\rvecxi}{\bmxi}
\safemath{\rveclambda}{\bmlambda}
\safemath{\rvecmu}{\bmmu}
\safemath{\rvectheta}{\bmtheta}
\safemath{\rvecphi}{\bmphi}

\safemath{\rmatA}{\bimA}
\safemath{\rmatB}{\bimB}
\safemath{\rmatC}{\bimC}
\safemath{\rmatD}{\bimD}
\safemath{\rmatE}{\bimE}
\safemath{\rmatF}{\bimF}
\safemath{\rmatG}{\bimG}
\safemath{\rmatH}{\bimH}
\safemath{\rmatI}{\bimI}
\safemath{\rmatJ}{\bimJ}
\safemath{\rmatK}{\bimK}
\safemath{\rmatL}{\bimL}
\safemath{\rmatM}{\bimM}
\safemath{\rmatN}{\bimN}
\safemath{\rmatO}{\bimO}
\safemath{\rmatP}{\bimP}
\safemath{\rmatQ}{\bimQ}
\safemath{\rmatR}{\bimR}
\safemath{\rmatS}{\bimS}
\safemath{\rmatT}{\bimT}
\safemath{\rmatU}{\bimU}
\safemath{\rmatV}{\bimV}
\safemath{\rmatW}{\bimW}
\safemath{\rmatX}{\bimX}
\safemath{\rmatY}{\bimY}
\safemath{\rmatZ}{\bimZ}

\safemath{\rmatDelta}{\bimDelta}
\safemath{\rmatLambda}{\bimLambda}
\safemath{\rmatPhi}{\bimPhi}
\safemath{\rmatSigma}{\bimSigma}
\safemath{\rmatOmega}{\bimOmega}
\safemath{\rmatTheta}{\bimTheta}

\usepackage{amssymb}
\usepackage{amsfonts}
\usepackage{mathrsfs}
\usepackage{xspace}
\usepackage{bm}
\usepackage{fancyref}
\usepackage{textcomp}

\usepackage{multirow}
\usepackage{stmaryrd}

\newenvironment{textbmatrix}{	\setlength{\arraycolsep}{2.5pt}%
								\big[\begin{matrix}}{\end{matrix}\big]%
								\raisebox{0.08ex}{\vphantom{M}}}

\def\be{\begin{equation}}
\def\ee{\end{equation}}
\def\een{\nonumber \end{equation}}
\def\mat{\begin{bmatrix}}
\def\emat{\end{bmatrix}}
\def\btm{\begin{textbmatrix}}
\def\etm{\end{textbmatrix}}

\def\ba#1\ea{\begin{align}#1\end{align}}
\def\bas#1\eas{\begin{align*}#1\end{align*}}
\def\bs#1\es{\begin{split}#1\end{split}}
\def\bg#1\eg{\begin{gather}#1\end{gather}}
\def\bml#1\eml{\begin{multline}#1\end{multline}}
\def\bi#1\ei{\begin{itemize}#1\end{itemize}}

\newcommand{\lefto}{\mathopen{}\left}

\DeclareMathOperator*{\argmin}{arg\;min}		
\DeclareMathOperator{\Exop}{\opE}			

\newcommand{\Ex}[2]{\ensuremath{\Exop_{#1}\lefto[#2\right]}} 	
\newcommand{\abs}[1]{\lefto\lvert#1\right\rvert}		

\newcommand{\vecnorm}[1]{\lefto\lVert#1\right\rVert}		
\newcommand{\frobnorm}[1]{\vecnorm{#1}_{\text{F}}}	

\safemath{\dirac}{\delta}					
\safemath{\krond}{\dirac}					

\safemath{\upto}{\uparrow}
\safemath{\downto}{\downarrow}
\safemath{\iu}{j}							
\safemath{\ev}{\lambda}						
\safemath{\hilseqspace}{l^{2}}				
\newcommand{\banachfunspace}[1]{\setL^{#1}}	
\safemath{\hilfunspace}{\banachfunspace{2}}	

\safemath{\SNR}{\textit{SNR}} 				
\safemath{\PAR}{\textit{PAR}} 				
\safemath{\No}{N_0}							
\safemath{\Es}{E_s}							
\safemath{\Eb}{E_b}							
\safemath{\EbNo}{\frac{\Eb}{\No}}
\safemath{\EsNo}{\frac{\Es}{\No}}

\DeclareMathOperator{\CHop}{\ensuremath{\opH}} 
\safemath{\tvir}{\rndh_{\CHop}}				
\safemath{\tvtf}{\rndl_{\CHop}}				
\safemath{\spf}{\rnds_{\CHop}}				
\safemath{\bff}{H_{\CHop}}					

\safemath{\ircf}{r_{h}}						
\safemath{\tftvcf}{r_{s}}					
\safemath{\tfcf}{r_{l}}						
\safemath{\bfcf}{r_{H}}						

\safemath{\tcorr}{c_h}						
\safemath{\scf}{c_{s}}						
\safemath{\tfcorr}{c_{l}}					
\safemath{\fcorr}{c_{H}}						

\safemath{\mi}{I}							
\safemath{\capacity}{C}						

\safemath{\normal}{\mathcal{N}}			
\safemath{\jpg}{\mathcal{CN}}			
\safemath{\mchain}{\leftrightarrow}		

\safemath{\dB}{\,\mathrm{dB}}
\safemath{\dBm}{\,\mathrm{dBm}}
\safemath{\Hz}{\,\mathrm{Hz}}
\safemath{\kHz}{\,\mathrm{kHz}}
\safemath{\MHz}{\,\mathrm{MHz}}
\safemath{\GHz}{\,\mathrm{GHz}}
\safemath{\s}{\,\mathrm{s}}
\safemath{\ms}{\,\mathrm{ms}}
\safemath{\mus}{\,\mathrm{\text{\textmu}s}}
\safemath{\ns}{\,\mathrm{ns}}
\safemath{\ps}{\,\mathrm{ps}}
\safemath{\meter}{\,\mathrm{m}}
\safemath{\mm}{\,\mathrm{mm}}
\safemath{\cm}{\,\mathrm{cm}}
\safemath{\m}{\,\mathrm{m}}
\safemath{\W}{\,\mathrm{W}}
\safemath{\mW}{\, \mathrm{mW}}
\safemath{\J}{\,\mathrm{J}}
\safemath{\K}{\,\mathrm{K}}
\safemath{\bit}{\,\mathrm{bit}}
\safemath{\nat}{\,\mathrm{nat}}

\safemath{\define}{\triangleq}			

\safemath{\equivalent}{\sim}
\safemath{\distas}{\sim}					
\safemath{\sdiff}{\Delta}				

\safemath{\reals}{\mathbb{R}}
\safemath{\positivereals}{\reals_{+}}
\safemath{\integers}{\mathbb{Z}}
\safemath{\posint}{\integers_{+}}
\safemath{\naturals}{\mathbb{N}}
\safemath{\posnaturals}{\naturals_{+}}
\safemath{\complexset}{\mathbb{C}}
\safemath{\rationals}{\mathbb{Q}}

\newcommand*{\fancyrefapplabelprefix}{app}		
\newcommand*{\fancyrefthmlabelprefix}{thm}		
\newcommand*{\fancyreflemlabelprefix}{lem}		
\newcommand*{\fancyrefcorlabelprefix}{cor}		
\newcommand*{\fancyrefdeflabelprefix}{def}		
\newcommand*{\fancyrefproplabelprefix}{prop}		
\newcommand*{\fancyrefexmpllabelprefix}{exmpl}
\newcommand*{\fancyrefalglabelprefix}{alg}		
\newcommand*{\fancyreftbllabelprefix}{tbl}		

\frefformat{vario}{\fancyrefseclabelprefix}{Section~#1}
\frefformat{vario}{\fancyrefthmlabelprefix}{Theorem~#1}
\frefformat{vario}{\fancyreftbllabelprefix}{Table~#1}
\frefformat{vario}{\fancyreflemlabelprefix}{Lemma~#1}
\frefformat{vario}{\fancyrefcorlabelprefix}{Corollary~#1}
\frefformat{vario}{\fancyrefdeflabelprefix}{Definition~#1}
\frefformat{vario}{\fancyreffiglabelprefix}{Fig.~#1}
\frefformat{vario}{\fancyrefapplabelprefix}{Appendix~#1}
\frefformat{vario}{\fancyrefeqlabelprefix}{(#1)}
\frefformat{vario}{\fancyrefproplabelprefix}{Proposition~#1}
\frefformat{vario}{\fancyrefexmpllabelprefix}{Example~#1}
\frefformat{vario}{\fancyrefalglabelprefix}{Algorithm~#1}

\safemath{\dictab}{[\,\dicta\,\,\dictb\,]}

\safemath{\ysig}{\bmy}
\safemath{\ysighat}{\hat{\ysig}}
\safemath{\ysigdim}{M}
\safemath{\xsig}{\bmx}
\safemath{\xsigdim}{N}
\safemath{\nx}{n_x}
\safemath{\zsig}{\bmz}
\safemath{\zsigdim}{\ysigdim}
\safemath{\rsig}{\bmr}
\safemath{\Adict}{\bA}
\safemath{\Adicttilde}{\widetilde{\Adict}}
\safemath{\Adictdim}{\outputdim\times\xsigdim}
\safemath{\avec}{\bma}
\safemath{\avectilde}{\tilde{\avec}}
\safemath{\Bdict}{\bB}
\safemath{\Bdicttilde}{\widetilde{\Bdict}}
\safemath{\Cdict}{\bC}
\safemath{\cvec}{\bmc}
\safemath{\Ddict}{\bD}
\safemath{\Ddictdim}{\ysigdim\times\xsigdim}
\safemath{\dvec}{\bmd}
\safemath{\Ddicttilde}{\widetilde{\bD}}
\safemath{\Bonb}{\bB}
\safemath{\bvec}{\bmb}
\safemath{\Bonbdim}{\ysigdim\times\ysigdim}
\safemath{\noise}{\bmn}
\safemath{\noisedim}{\ysigim}
\safemath{\err}{\bme}
\safemath{\errdim}{\ysigdim}
\safemath{\errset}{\setE}
\safemath{\nerr}{n_e}
\safemath{\delop}{\bP_\errset}
\safemath{\delopc}{\bP_{{\errset}^c}}

\safemath{\cplxi}{\imath}
\safemath{\cplxj}{\jmath}

\safemath{\dict}{\matD}
\safemath{\inputdim}{N}		
\safemath{\outputdim}{M}		
\safemath{\sparsity}{S}	
\safemath{\inputdimA}{{N_a}}	
\safemath{\inputdimB}{{N_b}}	
\safemath{\elemA}{{n_a}}	
\safemath{\elemB}{{n_b}}	
\safemath{\resA}{\matR_a}	
\safemath{\resB}{\matR_b}	
\safemath{\subD}{\matS} 
\safemath{\subA}{\matS_a} 
\safemath{\subB}{\matS_b} 
\safemath{\dicta}{\matA} 	
\safemath{\dictb}{\matB} 	
\safemath{\hollowS}{H}
\safemath{\hollowA}{H_a}
\safemath{\hollowB}{H_b}
\safemath{\cross}{Z}
\safemath{\coh}{\mu_d}			
\safemath{\coha}{\mu_a}			
\safemath{\cohb}{\mu_b}			
\safemath{\mubs}{\nu}	
\safemath{\cohm}{\mu_m} 
\safemath{\dictset}{\setD}	
\safemath{\dictsetp}{\dictset(\coh,\coha,\cohb)}	
\safemath{\dictsetgen}{\dictset_\text{gen}}
\safemath{\dictsetgenp}{\dictsetgen(\coh)}
\safemath{\dictsetonb}{\dictset_\text{onb}}
\safemath{\dictsetonbp}{\dictsetonb(\coh)}

\safemath{\leftside}{U}
\safemath{\rightsideA}{R_a}
\safemath{\rightsideB}{R_b}

\safemath{\indexS}{\setI_S} 

\safemath{\na}{n_a}			
\safemath{\nb}{n_b}			
\safemath{\coeffa}{p_i}	
\safemath{\coeffb}{q_j}	
\safemath{\seta}{\setP}		
\safemath{\setb}{\setQ}     
\safemath{\setw}{\setW}	
\safemath{\setz}{\setZ}	
\safemath{\cola}{\veca}		
\safemath{\colb}{\vecb}		
\safemath{\cold}{\vecd}		
\safemath{\inputvec}{\vecx} 	
\safemath{\error}{\vece}	
\safemath{\noiseout}{\vecz} 	
\safemath{\inputvecel}{x}
\safemath{\inputveca}{\vecx_a}
\safemath{\inputvecb}{\vecx_b}
\safemath{\outputvec}{\vecy}	
\safemath{\lambdamin}{\lambda_{\mathrm{min}}}

\safemath{\elltwo}{\ell_2}
\safemath{\ellone}{\ell_1}
\safemath{\ellzero}{\ell_0}
\safemath{\ellinf}{\ell_\infty}
\safemath{\ellinftilde}{\ell_{\widetilde\infty}}
\safemath{\licard}{Z(\coh,\coha,\cohb)}
\safemath{\xsol}{\hat{x}}
\safemath{\xbord}{x_b}		
\safemath{\xstat}{x_s}		
\safemath{\xstatLone}{\tilde{x}_s}
\safemath{\order}{\mathcal{O}} 
\safemath{\scales}{\Theta} 
\safemath{\ones}{\mathbf{1}} 
\safemath{\zeroes}{\mathbf{0}} 
\safemath{\thlone}{\kappa(\coh,\cohb)} 
\safemath{\constoneA}{\delta} 
\safemath{\constoneB}{\epsilon} 
\safemath{\nlarge}{L}				   
\safemath{\sumlarge}{S_\nlarge}
\safemath{\maxlarger}{P_\nlarge}	   
\safemath{\Pzero}{\textrm{P0}}	
\safemath{\Pone}{\textrm{P1}}
\safemath{\vecfir}{\vecw}			 
\safemath{\vecsec}{\vecz}
\safemath{\elvecfir}{w}              
\safemath{\elvecsec}{z}				 
\safemath{\nlargefir}{n}
\safemath{\normout}{\gamma}
\safemath{\auxfun}{h}
\safemath{\supp}{\textrm{supp}}

\safemath{\indexa}{\ell}
\safemath{\indexb}{r}
\safemath{\indexc}{i}
\safemath{\indexd}{j}

\safemath{\project}{P}

\usepackage{framed}

\IEEEoverridecommandlockouts
\allowdisplaybreaks 

\safemath{\Hj}{\bmj}
\safemath{\bsj}{\bmw}
\safemath{\sj}{w}
\safemath{\Ej}{E_w}
\safemath{\proxg}{\text{prox}_g}
\safemath{\pma}{\text{pma}_\setS}
\safemath{\rE}{\rho_{\textsf{\tiny{E}}}}
\safemath{\rP}{\rho_{\textsf{\tiny{P}}}}

\newcommand{\norm}[1]{\left\lVert #1 \right\rVert}

\newcommand{\PR}[1]{\ensuremath{\!\left[#1\right]}}
\newcommand{\PC}[1]{\ensuremath{\!\left(#1\right)}}
\newcommand{\chav}[1]{\ensuremath{\!\left\{#1\right\}}}

\begin{document}

\title{An Optimization-Based User Scheduling Framework for Multiuser MIMO Systems}

\author{\IEEEauthorblockN{ Victoria Palhares and Christoph Studer\\
}
\thanks{VP is with Nokia Bell Labs in Stuttgart, Germany; CS is with ETH Zurich in Switzerland; email: victoria.palhares@nokia-bell-labs.com, studer@ethz.ch.}
\thanks{The work of VP and CS was supported in part by an ETH Research Grant.}
\thanks{
We thank Haochuan Song and Seyed Hadi Mirfarshbafan for discussions on forward-backward splitting and mmWave channels, respectively, and we thank Gian Marti and Sueda Taner for their comments on the paper. We also thank Remcom for providing a license for the Wireless InSite ray-tracing software.}
\thanks{A short version of this paper was presented at IEEE SPAWC 2022 \cite{Palhares2022}. The new contributions are as follows: (i) improved control over UE activity through inequality constraints, (ii) one new objective function, (iii) detailed gradients of all proposed objective functions, (iv) a detailed derivation of the Karush-Kuhn-Tucker conditions for the simplex projection with inequality constraints, and (v) a performance evaluation in a cell-free scenario.}
\thanks{The MATLAB code will be released on GitHub after the review process.}
}

\maketitle


\begin{abstract}
Resource allocation is a key factor in multiuser (MU) multiple-input multiple-output (MIMO) wireless systems to provide high quality of service to all user equipments (UEs). 
In congested scenarios, UE scheduling enables UEs to be distributed over time, frequency, or space in order to mitigate inter-UE interference. 
Many existing UE scheduling methods rely on greedy algorithms, which fail at treating the resource-allocation problem globally. 
In this work, we propose a UE scheduling framework for MU-MIMO wireless systems that approximately solves a nonconvex optimization problem that treats scheduling globally. 
Our UE scheduling framework determines subsets of UEs that should transmit simultaneously in a given resource slot and is flexible in the sense that it (i) supports a variety of objective functions (e.g., post-equalization mean squared error, capacity, and achievable sum rate) and (ii) enables precise control over the minimum and maximum number of resources the UEs should occupy. 
We demonstrate the efficacy of our UE scheduling framework for millimeter-wave massive MU-MIMO and sub-6-GHz cell-free massive MU-MIMO systems, and we show that it outperforms existing scheduling algorithms while approaching the performance of an exhaustive search.
\end{abstract}

\begin{IEEEkeywords}
Cell-free, massive multiuser (MU) multiple-input multiple-output (MIMO), millimeter-wave (mmWave), optimization, resource allocation, scheduling.
\end{IEEEkeywords}

\section{Introduction}
Millimeter-wave (mmWave) and cell-free massive multiuser (MU) multiple-input multiple-output (MIMO) wireless systems are promising key technologies in next-generation wireless systems \cite{Swindlehurst2014,Zhang2020}. 
The combination of mmWave with massive MU-MIMO not only promises large beamforming gains to combat the high path loss at mmWave frequencies, but also enables high-bandwidth data transmission to multiple user equipments (UEs) in the same time-frequency resource. 
In cell-free systems, improved coverage compared to conventional cellular networks can be achieved by densely deploying access points (APs) that perform joint baseband processing \cite{Ngo2017}.

In order to maximize the quality of service (QoS) for all UEs in a network, resource allocation strategies, such as power control and UE scheduling,  are necessary. 
Power control mitigates the near-far problem between the UEs transmitting to a basestation (BS). 
UE scheduling distributes the UEs' requests in time, frequency, and/or space, to minimize inter-UE interference in scenarios where multiple UEs have similar channel impulse responses---this can happen in congested scenarios with many UEs transmitting in close vicinity. 

The UE scheduling literature widely focuses on greedy algorithms, where one UE is selected at a time to join a set of scheduled UEs \cite{Choi2019,Cui2018,He2017,Kim2018,Lee2018b,Paul2020,Wu2017,Zhu2021,Zhao2018,Xu2018,Zhang2022,Yoo2006,Lee2018a,Mashdour2022,Mashdour2023,Riera2019,Liu2022,Ko2021,Gong2022}. 
Such approaches (i) have difficulties when scheduling UEs over multiple resource slots or when lower and upper limits on the number of UEs per resource block are imposed, (ii) are typically restricted to one specific cost function, and (iii) fail to consider the UE scheduling problem globally, which can lead to suboptimal QoS compared to an exhaustive search (ES) over all possible schedules.

\subsection{Contributions}
We propose a novel UE scheduling framework for MU-MIMO wireless systems that solves the UE scheduling problem globally, over all UEs and all time slots in a joint manner.
We start by formulating UE scheduling as a discrete optimization problem that supports a variety of cost functions, such as post-linear minimum mean squared error (LMMSE) equalization mean squared error (MSE), channel capacity, and post-LMMSE equalization sum of achievable rates. 
In addition, our framework enables precise control over the minimum and maximum number of resources each UE should occupy. 
In order to efficiently find approximate solutions to the discrete UE scheduling problem, we first relax the combinatorial problem into a nonconvex but smooth optimization problem, which we solve approximately using forward-backward splitting~(FBS)~\cite{Parikh2014}. 
To employ FBS, we provide the gradients for each of the presented objective functions as well as the proximal operator. 
In our framework, the proximal operator performs an orthogonal projection onto a constraint set, which we define as the intersection of two simplexes described by inequalities. 
The inequalities in our UE scheduling problem represent the minimum and maximum amount of resources that UEs should occupy. 
To derive the proximal operator that finds solutions lying at the intersection of these two simplexes, we propose an algorithm to calculate this projection based on its Karush-Kuhn-Tucker (KKT) conditions. 
Upon completion of FBS, we propose a quantization algorithm that enforces binary-valued solutions while ensuring the constraints to be satisfied. 
We use channel vectors for mmWave massive MU-MIMO and sub-6-GHz cell-free massive MU-MIMO systems from a commercial ray tracer to demonstrate the efficacy of our method.
Our results demonstrate that the proposed framework outperforms a range of baseline methods in terms of per-UE uncoded bit error rate (BER), per-UE hard-output mutual information (HMI), per-UE MSE, and per-UE achievable rate, often closely approaching the performance of an ES over all possible schedules.

\subsection{Notation}
Upper- and lower-case bold letters denote matrices and vectors, respectively. 
We use~$A_{i,j}$ for the element in the $i$th row and $j$th column of $\bA$, $\bma_j$ as the $j$th column of~$\bA$, and~$a_i$ as the $i$th element of vector $\bma$. 
We define $\bI_M$, $\boldsymbol{1}_{L \times M}$, and~$\boldsymbol{0}_{L \times M}$ as the $M \times M$ identity matrix, $L \times M$ all-ones matrix, and $L \times M$ all-zeros matrix, respectively. 
The superscript $(\cdot)^{\text{H}}$ indicates Hermitian transposition. 
A diagonal matrix with $\bma$ on the main diagonal is denoted by $\text{diag} \PC{\bma}$; the determinant of matrix $\bA$ is $\det\PC{\bA}$. 
The absolute value, Euclidean norm, and Frobenius norm are denoted by $|\cdot|$, $\vecnorm{\cdot}_2$, and $\vecnorm{\cdot}_\text{F}$, respectively.
Expectation is $\Ex{}{\cdot}$ and~$\overset{e}{\leq} $ is element-wise less-or-equal-to. 

\subsection{Paper Outline}
The rest of the paper is organized as follows. 
\fref{sec:priorart} reviews prior art.
\fref{sec:prerequisites} introduces the system model and the scheduling problem. 
\fref{sec:UE_framework} presents the scheduling framework. 
\fref{sec:cost_functions} lists objective functions and their gradients.  
\fref{sec:proximal_operator} details the orthogonal projection in the constraint set. 
\fref{sec:quantization_algorithm} describes the proposed quantization algorithm.
\fref{sec:simulation_results} shows simulation results for mmWave and cell-free massive MU-MIMO scenarios.
\fref{sec:limitations} discusses limitations and future directions.
\fref{sec:conclusions} concludes.

\section{Relevant Prior Art}  
\label{sec:priorart}
UE scheduling has been widely studied in massive MU-MIMO systems \cite{Farsaei2019,Yang2018,Meng2018,Jiang2018,Lee2018a,Hu2016,Yoo2006,Hsu2024,Feres2023,Ko2021}, mmWave communication \cite{Hosseini2021,Zou2021,Sha2022,Zhang2022,Gao2020,Zhao2018,Cui2018,He2017,Zhu2021,Xu2018,Wu2017,Kim2018,Chukhno2021,Lee2021,Paul2020,Lee2018b,Choi2019,Gallyas-Sanhueza2024,Peng2021}, cell-free networks \cite{Liu2022,Eddine2018,Riera2019,Denis2021,Chen2019,Mashdour2022,Ammar2022b,Ammar2022a,Mashdour2023,Gottsch2024,Gong2022}, and non-orthogonal multiple access (NOMA) \cite{Zhang2022,Fang2017,Cui2018,Zhu2021,Lee2018b}. 
This work focuses on UE scheduling in all-digital massive MU-MIMO, mmWave, and cell-free networks. 
In what follows, a ``UE set'' refers to a UE cluster, group, partition, or the ensemble of UEs assigned to a time slot.

\subsection{Greedy Scheduling Algorithms} 
In massive MU-MIMO systems, we highlight the greedy UE scheduling algorithms in \cite{Yoo2006,Lee2018a}. 
The work in \cite{Yoo2006} proposes the semiorthogonal user selection (SUS) method, forming UE sets with nearly orthogonal channels; the work \cite{Lee2018a} extends SUS to distributed MIMO.

In mmWave systems, we highlight channel structure-based scheduling (CSS), greedy maximum sum rate (greedy), and chordal distance-based scheduling (Chordal) \cite{Choi2019}. 
In CSS, UEs with the largest signal-to-interference-plus-noise ratios (SINRs) are first added to a set~$\setS$, then refined to $\setK$ via semi-orthogonality. 
Chordal separates UEs in the angular domain. 
Other greedy mmWave methods are proposed in \cite{Cui2018,He2017,Kim2018,Lee2018b,Paul2020,Wu2017,Zhu2021,Zhao2018,Xu2018,Zhang2022}.

In cell-free networks, we highlight the greedy approaches in \cite{Mashdour2022,Mashdour2023,Riera2019}, where \cite{Mashdour2022,Mashdour2023} schedule small UE subsets sequentially to maximize sum rate; the method proposed in~\cite{Riera2019} excludes ill-conditioned UEs.

Overall, all of the above methods iteratively select UEs to greedily optimize an objective function, often getting stuck in local optima. 
In contrast, our framework solves the UE scheduling problem globally, jointly determining which UEs should transmit in each time slot via a single optimization problem---this approach often results in excellent UE schedules and outperforms greedy methods (see~\fref{sec:simulation_results}).

\subsection{UE-to-Beam Association in mmWave}
In mmWave systems, the high directionality of wave propagation at mmWave frequencies implies that what many prior works call ``UE scheduling\footnote{UE scheduling can be used in different contexts, such as time scheduling, frequency scheduling, or scheduling of UEs to certain beams.}'' is in fact UE-to-beam association~\cite{Cui2018,He2017,Kim2018,Lee2018b,Gao2020,Paul2020,Wu2017,Hosseini2021,Lee2021,Zhu2021,Chukhno2021}. 
In contrast, we focus on UE-to-time-slot association. Our framework could be extended to UE-to-beam association, which we leave for future work.

\subsection{UE-Centric Cell-Free Networks}
In cell-free networks, each UE is typically close to only a few APs and far from others, leading to channel sparsity \cite{Song2021}. 
Prior work introduced UE-centric architectures \cite{Buzzi2017} exploiting this property, where UEs transmit only to selected APs. 
In~\cite{Ammar2022a,Ammar2022b}, the authors formulate a joint UE scheduling, power allocation, and beamforming problem for the downlink of a UE-centric cell-free massive MU-MIMO system. 
In this work, each UE is first assigned to specific APs, forming UE sets per AP, after which scheduling is performed independently at each AP, limited by its number of antennas. In \cite{Liu2022}, the authors propose a greedy joint AP-UE clustering strategy to maximize throughput in cell-free visible-light networks with both time and frequency scheduling. 
Other frequency-scheduling approaches include~\cite{Denis2021,Gong2022}, aiming to mitigate inter-UE set interference and maximize throughput, respectively.

In contrast, we consider a fully cooperative cell-free scenario in which all UEs communicate with all APs and scheduling is performed globally across all AP antennas and UEs. 
Furthermore, we focus on time scheduling, leaving frequency scheduling for future work.

\subsection{Other Scheduling Algorithms}
While most prior work relies on greedy algorithms, there are some machine learning (ML)-based exceptions \cite{Hsu2024,Feres2023,Chukhno2021,Zou2021}. In \cite{Hsu2024,Feres2023}, channel state information (CSI) is projected into low-dimensional embeddings and clustered based on similarity. 
In \cite{Chukhno2021}, a self-supervised neural network splits UEs receiving the same data into subsets served by the same beam according to azimuth similarity. In \cite{Zou2021}, UE classification via $k$-means clustering minimizes inter-UE interference.

As a low-complexity alternative to greedy and ML-based methods, reference~\cite{Gallyas-Sanhueza2024} proposes low-fidelity scheduling (LoFi, LoFi++). 
LoFi performs multiple random scheduling trials and selects the best based on a given objective, while LoFi++ improves this by reassigning the most critical UE to another set to test for performance gains, introducing greedy-like behavior.

Such ML-based and heuristic methods lack explicit and precise control over resource allocation, as they cannot enforce upper and lower bounds on resources per UE. 
In contrast, our framework provides precise control and flexibility, allowing, for example, specification of the number of time slots per UE and UEs per time slot.

\subsection{Resource Allocation}
As mentioned above, most scheduling algorithms in the massive MU-MIMO, mmWave, and cell-free literature lack flexibility in resource allocation. 
Many provide no guarantee that every UE belongs to at least one UE set \cite{Yoo2006,Choi2019,Kim2018,Gao2020,Zhao2018,Xu2018,Zhang2022,Mashdour2022,Riera2019,Hosseini2021,Farsaei2019,Jiang2018,Lee2018a,Chen2019,Hsu2024,Mashdour2023,Gottsch2024,Ko2021,Gong2022}. 
Furthermore, most methods are unable to enforce minimum and maximum numbers of UEs per set or sets per UE \cite{Yoo2006,Choi2019,Cui2018,He2017,Kim2018,Gallyas-Sanhueza2024,Lee2018b,Gao2020,Paul2020,Wu2017,Zhu2021,Zhao2018,Xu2018,Zhang2022,Mashdour2022,Riera2019,Hosseini2021,Chukhno2021,Zou2021,Lee2021,Farsaei2019,Yang2018,Meng2018,Jiang2018,Lee2018a,Hu2016,Liu2022,Eddine2018,Denis2021,Chen2019,Ammar2022b,Ammar2022a,Fang2017,Peng2021,Feres2023,Hsu2024,Mashdour2023,Gottsch2024,Ko2021,Gong2022}. 
Finally, some methods do not allow a UE to belong to multiple sets simultaneously (e.g., to be served in multiple time slots)\cite{Gallyas-Sanhueza2024,Cui2018,He2017,Lee2018b,Gao2020,Wu2017,Zhu2021,Zhang2022,Sha2022,Chukhno2021,Zou2021,Yang2018,Hu2016,Liu2022,Fang2017,Peng2021,Hsu2024,Feres2023,Gong2022}.

In contrast, our framework enforces minimum and maximum numbers of UEs per set and sets per UE, guarantees that every UE belongs to at least one set, and allows intersections between UE sets, allowing a UE to belong to multiple sets if~desired.

\subsection{Cost Functions}
Beyond resource allocation, many scheduling algorithms in the massive MU-MIMO, mmWave, and cell-free literature lack objective-function flexibility. 
Existing methods typically address only one aspect of the problem: some enforce UE orthogonality \cite{Yoo2006,Choi2019,Lee2018a}, while others optimize a specific metric, such as maximization of the SINR \cite{Choi2019,Gallyas-Sanhueza2024}, achievable rate \cite{Choi2019,Cui2018,He2017,Kim2018,Gao2020,Zhao2018,Mashdour2022,Riera2019,Farsaei2019,Jiang2018,Liu2022,Chen2019,Ammar2022b,Ammar2022a,Hsu2024,Feres2023,Mashdour2023,Gottsch2024,Gong2022}, energy efficiency (EE) \cite{Lee2018b,Zhu2021}, spectral efficiency (SE) \cite{Zhu2021,Yang2018,Zhang2022}, channel capacity \cite{Hu2016}, channel energy \cite{Wu2017,Xu2018,Fang2017}, or minimization of the interference, BER, and channel correlation \cite{Paul2020,Denis2021,Ko2021,Zou2021,Meng2018,Eddine2018,Peng2021}.

In contrast, our optimization-based UE scheduling framework is general as it supports any differentiable cost functions that can be expressed in terms of the UE set. 
To demonstrate flexibility, we propose a post-equalization MSE objective function based on the LMMSE equalizer, alongside the conventional channel capacity and achievable rate objective functions.

\subsection{All-Digital Basestation Architectures}
Beyond their reliance on greedy algorithms, most mmWave scheduling works focus on hybrid analog-digital \cite{Choi2019,He2017,Kim2018,Gao2020,Paul2020,Wu2017,Hosseini2021,Zhu2021,Zhao2018,Xu2018,Zhang2022,Sha2022,Chukhno2021,Zou2021,Lee2021,Peng2021} or fully analog \cite{Cui2018,Lee2018b} beamforming, where each BS has fewer data converters than antennas and performs part of the beamforming in the analog domain.

In contrast, we consider fully digital BSs and APs, in which each antenna is equipped with a pair of analog-to-digital converters that sample the in-phase and quadrature baseband signals. 
Such architectures can match the energy efficiency of hybrid systems \cite{Dutta2020,Roth2018,Skrimponis2020} while simplifying baseband processing tasks (e.g.,  channel estimation, data detection, MU beamforming, and UE scheduling).


\section{Prerequisites} \label{sec:prerequisites}

\subsection{System Model} \label{sec:system_model}
We consider the uplink of an all-digital massive MU-MIMO system in which $U$ single-antenna UEs transmit data to an all-digital BS with $B$ receive antennas, each with its dedicated pair of data converters. 
Depending on the scenario, the receive antennas are deployed differently. 
In the mmWave scenario, we consider one receiving BS ($L=1$) with an $n_{\text{A}}$-antenna uniform linear array (ULA). 
In the sub-6-GHz cell-free scenario, we consider $L$ APs each with an $n_{\text{A}}$-antenna ULA.
In both cases, $B = Ln_{\text{A}}$ is the number of receive antennas. 
We consider a block-fading scenario with $t = 1,\dots,T$ time slots, frequency-flat channels\footnote{Our framework can be applied to frequency-selective channels via orthogonal frequency-division multiplexing (OFDM) or orthogonal frequency-division multiple access (OFDMA), which also enables frequency scheduling.}, and the standard input-output relation:
\begin{align}
\label{eq:received_vector}
\bmy \PR{t} = \bH \bms\PR{t} + \bmn \PR{t}\!.
\end{align}
Here,  $\bmy \PR{t}\in\opC^{B}$ is the receive vector in time slot $t$, $\bms\PR{t}  \in \opC^{U}$ contains the transmit symbols of all $U$ UEs, and $\bmn\PR{t} \in \opC^{B}$ models noise whose entries are i.i.d.\ circularly-symmetric complex Gaussian with variance $N_0$. 
The effective channel matrix $\bH= \sqrt{\frac{\rho_u}{\No}} \ \overline{\bH} \matDelta$  in \fref{eq:received_vector} combines the effect of the UE transmit power~$\rho_u$, the noise power $N_0$, the uplink MIMO channel matrix $\overline{\bH} \in \opC^{B \times U}$, and the power control matrix $\matDelta = \text{diag} \PC{\delta_1,\dots,\delta_U}$, whose entries are given by~\cite{Song2021} 
\begin{align} \label{eq:power_coefficients}
\delta_u^2 &= \frac{1}{\|\overline{\bmh}_u\|_2^2}\text{min} \chav{\| \overline{\bmh}_u\|_2^2,10^{\frac{\eta}{10}} \text{min}_{u'=1,\dots,U} \|\overline{\bmh}_{u'}\|_2^2}. 
\end{align}
Here, $\overline{\bmh}_u$ denotes the $u$th column of $\overline{\bH}$ and $\eta\geq0$ determines the maximum dynamic range between the weakest and strongest UE receive power in decibels (dB). 

In what follows, we assume orthogonal channel estimation, i.e., the UEs transmit orthogonal pilots in a dedicated training phase. 
With this training scheme, we can model the channel estimation matrix as $\hat{\bH} = \bH+\bE$, where $\bE \in \opC^{B \times U}$ is a noise matrix with i.i.d.\ circularly-symmetric complex Gaussian entries and variance $\frac{N_0}{U}$. 
On top of this training scheme, for the channel vectors obtained from a mmWave massive MU-MIMO system, we use the BEACHES algorithm \cite{Mirfarshbafan2020}, which performs denoising independently on each column of $\hat{\bH}$. 
For the channel vectors obtained from a sub-6-GHz cell-free massive MU-MIMO system, we simply use least squares channel estimation.

\subsection{UE Scheduling}
To formalize the UE scheduling problem, we define a binary-valued UE scheduling matrix $\bC \in \chav{0,1}^{U \times T}$, where $C_{u,t}=1$ and  $C_{u,t}=0$ indicate that UE $u$ during time slot $t$ is active and inactive, respectively. 
Furthermore, we define a diagonal mask matrix $\bD_{\bC}\PR{t}= \text{diag}\PC{\bmc_t} \in \chav{0,1}^{U \times U}$, where~$\bmc_t$ is the $t$th column of $\bC$. 
Through multiplication with~$\bms\PR{t}$, this mask matrix describes which UEs transmit symbols and which UEs are idle in time slot $t$. 

Our goal is to develop a framework that determines a UE scheduling matrix $\bC$ by minimizing a given objective function~$F \PC{\bC}\in\reals$ subject to constraints that specify the minimum and maximum number of resources that UEs are allowed to occupy.
We absorb the effect of the mask matrix in the effective channel matrix, which now depends on $t$, as 
\begin{align}
\bH\PR{t} = \bH \bD_{\bC}\PR{t} = \hat{\bH}\bD_{\bC}\PR{t} - \bE \bD_{\bC}\PR{t}\!.
\end{align} 
Here, $\hat{\bH}\PR{t} = \hat{\bH} \bD_{\bC}\PR{t}$ and $\bE\PR{t} = \bE \bD_{\bC}\PR{t}$;
both of these quantities will be needed for the computation of certain objective functions and gradients described in \fref{sec:cost_functions} and the performance metrics described in \fref{sec:performance_metrics}.

\section{UE Scheduling Framework}\label{sec:UE_framework}

\subsection{UE Scheduling as an Optimization Problem}
 The proposed UE scheduling framework requires an application-specific objective function $F\PC{\bC}$ and a set of constraints that determine the resource utilization in time and space. 
With this information, we wish to solve the following optimization problem:
\begin{align}   \label{eq:original_problem}
\underset{\bC \in \, \chav{0,1}^{U \times T}}{\text{minimize}} \  F\PC{\bC} \  \text{subject to} \ \bC \in \setC_U  \cap \setC_T,
\end{align}
with the two constraint sets 
 \begin{align} 
 \setC_U &= \big\{U_{\text{min}} \boldsymbol{1}_{1 \times T} \overset{e}{\leq} \boldsymbol{1}_{1 \times U}\bC \overset{e}{\leq} U_{\text{max}} \boldsymbol{1}_{1 \times T}\big\} \textnormal{, and} \label{eq:set_1} \\
 \setC_T  &= \big\{T_{\text{min}} \boldsymbol{1}_{U \times 1} \overset{e}{\leq} \bC\boldsymbol{1}_{T \times 1} \overset{e}{\leq} T_{\text{max}}  \boldsymbol{1}_{U \times 1}\big\}.  \label{eq:set_2}
\end{align}
The set $\setC_U$ in \fref{eq:set_1} determines the minimum $U_{\text{min}}$ and maximum~$U_{\text{max}}$ number of UEs allowed to transmit simultaneously per time slot; 
the set $\setC_T$ in \fref{eq:set_2} determines the minimum~$T_{\text{min}}$ and maximum~$T_{\text{max}}$ number of time slots each UE is allowed to transmit. 
Due to the discrete nature of the UE scheduling matrix $\bC$, the optimization problem in \fref{eq:original_problem} is combinatorial. 
For example, in a scenario with $U=32$ UEs where $16$ UEs transmit in the first time slot and the remaining $16$ UEs in the second time slot, an ES would need to test over $600$\,M scheduling matrices. 
Evidently, computationally efficient methods that approximately solve the problem in \fref{eq:original_problem} are necessary. 

\subsection{Problem Relaxation} \label{sec:problem_relaxation}
To arrive at an efficient UE scheduling algorithm, we relax the binary-valued $\{0,1\}$ entries in $\bC$ to the continuous range~$[0,1]$, i.e., $\bC \in \PR{0,1} ^{U \times T}$. 
Although this relaxation enables the use of computationally efficient gradient descent-based methods, it no longer enforces that the solutions are in~$\{0,1\}$. 
To mitigate this issue, we augment the objective function of the relaxed optimization problem with a regularizer~$R\PC{\bC}$ that promotes binary-valued entries in~$\bC$ (see \fref{sec:regularizer}). 
Taking the regularizer into account, we modify~\fref{eq:original_problem} into the following relaxed optimization problem:
\begin{align}   \label{eq:relaxed_problem}
\underset{\bC \in \, \PR{0,1}^{U \times T}}{\textnormal{minimize}} \  F\PC{\bC}  +  R\PC{\bC} \  \text{subject to } \ \bC \in \setC_U  \cap \setC_T.
\end{align}
Using $F\PC{\bC}  +  R\PC{\bC}$ as the new objective function, we are now able to deploy numerical optimization methods to efficiently compute binary-valued solutions to the relaxed problem in~\fref{eq:relaxed_problem}. 
Since the augmented objective function is, in general, nonconvex (even if the original objective function $F(\bC)$ is convex in the relaxed domain), numerical methods may only converge to a local minimum. 
Thus, to improve the quality of the solution, one can perform $K_{\text{init}}$ random initializations (see \fref{tbl:mmwave_and_cell_free_scenario}) and use the solution candidate~$\bC^*$ with the lowest cost $F\PC{\bC^*}$.

\subsection{Approximate Scheduling via Forward-Backward Splitting} \label{sec:fbs}
To approximately solve \fref{eq:relaxed_problem} for the set $\bC \in \PR{0,1} ^{U \times T}$, we use FBS~\cite{Goldstein2014}. 
This method performs the following step for iterations $i=1,2,\ldots,I_{\text{max}}$:
\begin{align} \label{eq:gradient_and_proximal}
\bC^{(i+1)} =\!\!\!\underset{\setX_{\setC_U\cap\setC_T}}{\mathrm{prox}} \!\Bigl(\bC^{(i)} - \tau^{(i)} \bigl(\nabla F(\bC^{(i)})+\nabla R(\bC^{(i)})\bigl)\!\Bigl).
\end{align}
Here, $\mathrm{prox}_{\setX_{\setC_U\cap\setC_T}}(\cdot)$ is the proximal operator of the indicator function of the set $\setC_U\cap \setC_T$, which corresponds to the orthogonal projection onto the set $\setC_U  \cap \setC_T $. 
The indicator function is defined  as follows:
\begin{align} \label{eq:indicator_function}
\setX_{\setC} \PC{\bX} = 
\left\{\begin{array}{ll}
0, & \text{ for }\bX \in \setC,\\
\infty, & \text{otherwise.}
\end{array}\right.
\end{align}
The proximal operator is detailed in \fref{sec:proximal_operator}.
The step size~$\tau^{(i)}$ of the $i$th iteration is selected
as detailed in~\cite{Goldstein2014,Zhou2006}. 
Also in \fref{eq:gradient_and_proximal}, we have $\nabla F(\cdot)$ and $\nabla R(\cdot)$, which are the gradients of $F\PC{\cdot}$ and  $R\PC{\cdot}$, respectively, and are described in further detail in \fref{sec:cost_functions}.
FBS is initialized with a matrix~$\bC^{(1)} \in \PR{0,1} ^{U \times T}$ drawn uniformly at random. 
The entries of the matrix in the last iteration $\bC^{(I_\text{max})}$ are quantized to binary values while satisfying the constraint sets $\setC_U$ and $\setC_T$; see \fref{sec:quantization_algorithm} for the details. 
The purpose of quantizing the entries of $\bC^{(I_\text{max})}$ to binary values is to determine whether a UE $u$ in time slot $t$ is active or inactive, i.e., whether $C^{(I_\text{max})}_{u,t}$ is equal to $1$ or $0$. 


\section{Objective Functions and Gradients} \label{sec:cost_functions}
To showcase the flexibility of our UE scheduling framework, we now present three options for $F\PC{\bC}$ as well as their gradients $\nabla F(\bC)$. 
We also detail the regularizer $R\PC{\bC}$ that promotes binary-valued solutions and its gradient $\nabla R(\bC)$. 
Note that the gradients $\nabla F(\bC) \in \opR^{U \times T}$ and $\nabla R(\bC) \in \opR^{U \times T}$ are the matrices of partial derivatives with elements $\frac{\partial F(\bC) }{\partial C_{u,t}}$ and $\frac{\partial R(\bC) }{\partial C_{u,t}}$, respectively, where $u=1,\dots,U$ and $t=1,\dots,T$.

\subsection{Post-LMMSE Equalization MSE} \label{sec:lmmse_mse}
 \subsubsection{Objective Function} \label{sec:MSE_LMMSE_cost_function}
For the first objective function, we consider the post-LMMSE equalization MSE, which is the MSE between the original transmit symbol vector $\bms\PR{t}$ and the estimated transmit symbol vector $\hat\bms\PR{t} = \bW\PR{t}^{\text{H}} \bmy\PR{t}$, where $\bW\PR{t}$ is an LMMSE equalizer. 
We define this objective function~as
\begin{align} \label{eq:mse_lmmse_cost_function}
F(\bC) = \sum_{t=1}^T \Ex{}{\vecnorm{\bD_{\bC}\PR{t}\bms\PR{t}- \bD_{\bC}\PR{t}\bW\PR{t}^{\text{H}} \bmy\PR{t}}_2^2}\!,
\end{align}
where expectation is taken over $\bms\PR{t}$ and $\bmn\PR{t}$ and\footnote{In our simulations, we reduce the complexity of the LMMSE equalizer calculation by doing a $U \times U$ inversion. However, to simplify the presentation of our derivations, we have kept \fref{eq:lmmse_equalizer} in its original form.} 
\begin{align} \label{eq:lmmse_equalizer}
\bW\PR{t} = \PC{\hat{\bH}\bD_{\bC}\PR{t}^2\hat{\bH}^{\text{H}}+ N_0\bI_B}^{-1}\hat{\bH}\bD_{\bC}\PR{t}
\end{align}
is the LMMSE equalization matrix in time slot $t$.

\subsubsection{Gradient} \label{sec:MSE_LMMSE_gradient}
The partial derivatives of the post-LMMSE equalization MSE objective function in \fref{eq:mse_lmmse_cost_function} with respect to an entry of the scheduling matrix $\bC$ are given by
\begin{align}
    \frac{\partial F(\bC) }{\partial C_{i,t}} = &2 C_{i,t}\big(1 + 
     \hat{\bmh}_i^{\text{H}}\bL_1\PR{t}\bL_2\PR{t} \bL_1\PR{t}\hat{\bmh}_i \nonumber\\
     & \qquad \,\, - 2C_{i,t}^2 \hat{\bmh}_i^{\text{H}}\bL_1\PR{t}\hat{\bmh}_i \big),
\end{align}
where $\hat{\bmh}_i \in \opC^{B \times 1}$ is the $i$th column of the matrix $\hat{\bH}$,
\begin{align} \label{eq:l1_t}
\bL_1\PR{t} = \PC{\hat{\bH}\bD_{\bC}\PR{t}^2\hat{\bH}^{\text{H}}+ N_0\bI_B}^{-1},
\end{align}
and $\bL_2\PR{t} = \hat{\bH}\bD_{\bC}\PR{t}^4\hat{\bH}^{\text{H}}$. \footnote{In the above expressions, the exponent in $\bD_{\bC}\PR{t}$ can be omitted as the entries are in $\{0,1\}$; this simplification has a marginal performance impact.} 

\subsection{Channel Capacity}
\subsubsection{Objective Function} \label{sec:capacity_cost_function}
For the second objective function, we consider the channel capacity. For simplicity, we take the capacity of the estimated channel $\hat{\bH}\PR{t}$ and treat it as the one from the real channel $\bH\PR{t}$. 
We define this objective function~as
\begin{align} \label{eq:capacity_cost_function}
        F(\bC)  =  -\sum_{t=1}^T \log_2 \PC{\det\PC{\bI_B + \frac{1}{N_0} \hat{\bH}\bD_{\bC}\PR{t}^2\hat{\bH}^{\text{H}}}}\!.
    \end{align}
Note that this objective function is the negative of the channel capacity, as the goal is to maximize this quantity.
    
\subsubsection{Gradient} \label{sec:capacity_gradient}
The partial derivative of the channel capacity objective function in \fref{eq:capacity_cost_function} with respect to one entry of the scheduling matrix $\bC$ is given by
\begin{align}
    \frac{\partial F(\bC) }{\partial C_{i,t}} = -2 \frac{1}{N_0} C_{i,t}\PC{\log_2{e}} \hat{\bmh}_i^{\text{H}} \bL_3\PR{t}^{-1} \hat{\bmh}_i,
\end{align}
where $e$ is Euler's number and $\bL_3\PR{t} = \bI_B + \frac{1}{N0} \hat{\bH}\bD_{\bC}\PR{t}^2\hat{\bH}^{\text{H}}$.

\subsection{Post-LMMSE Equalization Sum of Achievable Rates}  \label{sec:lmmse_sum_rate}
\subsubsection{Objective Function} \label{sec:rate_cost_function}
For the third objective function,  we consider the post-LMMSE equalization sum of achievable rates, which is the sum of achievable rates over all UEs and time slots, after the receive vector $\bmy\PR{t}$ goes through the LMMSE equalizer $\bW\PR{t}$ in \fref{eq:lmmse_equalizer}. 
Similarly to the objective function in~\fref{eq:capacity_cost_function}, we want to maximize the sum of achievable rates, and thus, we minimize the negative of this expression. We define this objective function as
\begin{align} \label{eq:sum_rate_cost_function}
F(\bC) = -\sum_{t=1}^T \sum_{u=1}^U \log_2\PC{1+\textit{SINR}_u\PC{\bmc_t} }\!,
\end{align}
where the post-equalization SINR of the $u$th UE in time slot $t$ is given by $ \textit{SINR}_u\PC{\bmc_t} =  \frac{\phi\PC{\bmc_t}}{\Psi\PC{\bmc_t}}$.
In this expression, we define
\begin{align}
\phi\PC{\bmc_t} &= \abs{\bmw_u\PR{t}^{\text{H}} \hat{\bmh}_u\PR{t}}^2 ,
\label{eq:des_signal_power}
\end{align}
which is the desired signal power and the denominator
\begin{align} \label{eq:sinr_gradient_denominator}
\Psi\PC{\bmc_t} &= \varphi\PC{\bmc_t} + \varrho\PC{\bmc_t} + \Phi\PC{\bmc_t}.
\end{align}
In \fref{eq:sinr_gradient_denominator}, we define interference, channel estimation error, and noise powers as follows:
\begin{align}
\varphi\PC{\bmc_t} &= \sum_{j=1,j\neq u}^U \abs{\bmw_u\PR{t}^{\text{H}} \!\hat{\bmh}_{j}\PR{t}}^2 \!\! \label{eq:interf_power}, \\
\varrho\PC{\bmc_t} &= \vecnorm{\bmw_u\PR{t}}_2^2 \sum_{j=1}^U C^2_{j,t} \Gamma_j \label{eq:channel_est_power}, \text{ and }\\
\Phi\PC{\bmc_t}  &= N_0 \vecnorm{\bmw_u\PR{t}}_2^2. \label{eq:noise_power}
\end{align}

\setcounter{equation}{23}
\begin{figure*}[!t]
    \begin{align}
    \frac{\partial \phi\PC{\bmc_t}}{\partial C_{i,t}}  = 
    \begin{cases} 
    4 C_{u,t}^3 \PC{\hat{\bmh}_u^{\text{H}}  \bL_1\PR{t}  \hat{\bmh}_u}^2 \big(1- C_{u,t}^2 \PC{\hat{\bmh}_u^{\text{H}}  \bL_1\PR{t}  \hat{\bmh}_u}\big), & \text{for}\ i = u,\\
    -4 C_{u,t}^4 C_{i,t} \PC{\hat{\bmh}_u^{\text{H}}  \bL_1\PR{t}  \hat{\bmh}_u}\PC{\hat{\bmh}_i^{\text{H}}\bL_1\PR{t} \hat{\bmh}_u \hat{\bmh}_u^{\text{H}} \bL_1\PR{t} \hat{\bmh}_i}, & \text{for}\  i \neq u.
    \label{eq:gradient_desired_signal}
    \end{cases}
    \end{align}

    \begin{align}
    \frac{\partial \varphi\PC{\bmc_t}}{\partial C_{i,t}} = \!\!
    \begin{cases}
    \sum_{j=1,j\neq u}^U 2C_{u,t} C_{j,t}^2\big(\hat{\bmh}_j^{\text{H}}\bL_1\PR{t}  \hat{\bmh}_u \hat{\bmh}_u^{\text{H}} \bL_1\PR{t}  \hat{\bmh}_j - 2C_{u,t}^2 \hat{\bmh}_j^{\text{H}}\PC{\bL_1\PR{t} \hat{\bmh}_u \hat{\bmh}_u^{\text{H}}}^2 \bL_1\PR{t} \hat{\bmh}_j \big), \\ \text{for}\ i = u,\\
    \sum_{j=1,j\neq u}^U 2C_{j,t} C_{u,t}^2 \big(\hat{\bmh}_j^{\text{H}}\bL_1\PR{t}  \hat{\bmh}_u \hat{\bmh}_u^{\text{H}} \bL_1\PR{t}  \hat{\bmh}_j - 2C_{j,t}^2 \hat{\bmh}_u^{\text{H}}\PC{\bL_1\PR{t} \hat{\bmh}_j \hat{\bmh}_j^{\text{H}}}^2 \bL_1\PR{t} \hat{\bmh}_u \big), \\ \text{for}\  i = j,\\
    -2 C_{u,t}^2 C_{j,t}^2 C_{i,t} \big(\hat{\bmh}_j^{\text{H}}\bL_1\PR{t} \hat{\bmh}_i \hat{\bmh}_i^{\text{H}}\bL_1\PR{t} \hat{\bmh}_u  \hat{\bmh}_u^{\text{H}}\bL_1\PR{t} \hat{\bmh}_j + \hat{\bmh}_j^{\text{H}} \bL_1\PR{t} \hat{\bmh}_u \hat{\bmh}_u^{\text{H}}\bL_1\PR{t} \hat{\bmh}_i \hat{\bmh}_i^{\text{H}}\bL_1\PR{t} \hat{\bmh}_j \big) \text{, for}\  i \neq u, i \neq j.
    \label{eq:gradient_interference}
    \end{cases}
    \end{align}
    
    \begin{align}
    \frac{\partial \varrho\PC{\bmc_t}}{\partial C_{i,t}}  = 
    \begin{cases}
    \PC{2 C_{u,t} \PR{\hat{\bmh}_u^{\text{H}} \bL_1\PR{t}\bL_1\PR{t}\hat{\bmh}_u - 2C_{u,t}^2 \hat{\bmh}_u^{\text{H}} \bL_1\PR{t}\bL_1\PR{t}\hat{\bmh}_u \hat{\bmh}_u^{\text{H}}\bL_1\PR{t} \hat{\bmh}_u}\bigg)} \\
    \bigg(\sum_{j=1}^U C_{j,t}^2 \Gamma_j \bigg) + \PC{\vecnorm{\bmw_u\PR{t}}_2^2} \PC{2 C_{u,t} \Gamma_u} \text{, for}\ i = u,\\
    \PC{- 2 C_{u,t}^2 C_{i,t} \PC{\hat{\bmh}_i^{\text{H}}\bL_1\PR{t}
    \bL_1\PR{t}\hat{\bmh}_u \hat{\bmh}_u^{\text{H}}\bL_1\PR{t}\hat{\bmh}_i + \hat{\bmh}_u^{\text{H}}\bL_1\PR{t}\bL_1\PR{t}\hat{\bmh}_i \hat{\bmh}_i^{\text{H}}\bL_1\PR{t}\hat{\bmh}_u}} \bigg(\sum_{j=1}^U C_{j,t}^2 \Gamma_j \bigg)\\
    + \PC{\vecnorm{\bmw_u\PR{t}}_2^2} \PC{2 C_{i,t} \Gamma_i} \text{, for}\  i \neq u.
    \label{eq:gradient_chest}
    \end{cases}
    \end{align}

    \begin{align} 
    \frac{\partial \Phi\PC{\bmc_t} }{\partial C_{i,t}}  = 
    \begin{cases}
    2 N_0 C_{u,t} \big(\hat{\bmh}_u^{\text{H}} \bL_1\PR{t} \bL_1\PR{t} \hat{\bmh}_u - 2C_{u,t}^2 \hat{\bmh}_u^{\text{H}}\bL_1\PR{t} \bL_1\PR{t} \hat{\bmh}_u \hat{\bmh}_u^{\text{H}}\bL_1\PR{t} \hat{\bmh}_u \big) \text{, for}\ i = u,\\
    - 2 N_0 C_{u,t}^2 C_{i,t} \big(\hat{\bmh}_i^{\text{H}}\bL_1\PR{t} \bL_1\PR{t} \hat{\bmh}_u \hat{\bmh}_u^{\text{H}} \bL_1\PR{t} \hat{\bmh}_i + \hat{\bmh}_u^{\text{H}}\bL_1\PR{t} \bL_1\PR{t} \hat{\bmh}_i \hat{\bmh}_i^{\text{H}} \bL_1\PR{t} \hat{\bmh}_u \big) \text{, for}\  i \neq u.
    \label{eq:gradient_noise}
    \end{cases}
    \end{align}
    \hrule
\end{figure*}

In the above expressions, $\bmw_u\PR{t}$ is the $u$th column of~$\bW\PR{t}$ from \fref{eq:lmmse_equalizer}, and $\hat{\bmh}_u\PR{t}$ is the $u$th column of the channel estimation matrix $\hat{\bH}\PR{t} = \hat{\bH} \bD_{\bC}\PR{t}$. 
Additionally, in \fref{eq:channel_est_power}, $\Ex{}{\bme_u\bme_u^{\text{H}}} = \Gamma_u \, \bI_B$, for $u=1,\dots,U$, where $\bme_u$ is the $u$th column of the channel estimation error matrix $\bE$.
For the mmWave massive MU-MIMO scenario, we approximate $\Gamma_u$ to the per-UE MSE output of the BEACHES algorithm~\cite{Mirfarshbafan2020} mentioned in \fref{sec:system_model}. 
For the sub-6-GHz cell-free massive MU-MIMO scenario, $\Gamma_u = N_0/U$. 
We highlight that $\textit{SINR}_u\PC{\bmc_t}$ contains $\bmc_t$ because $\bW\PR{t}$ and $\bH\PR{t}$ depend on $\bmc_t$.

\subsubsection{Gradient} \label{sec:rate_gradient}

The partial derivatives of the post-LMMSE equalization sum of achievable rates objective in \fref{eq:sum_rate_cost_function} with respect to one entry of the scheduling matrix $\bC$ are given by
\setcounter{equation}{21}
\begin{align}
	&\frac{\partial F(\bC)}{\partial C_{i,t}}\!=\!-\sum_{t=1}^T\!\sum_{u=1}^U  \!\frac{\PC{\log_2{e}}}{\PC{1+\textit{SINR}_u\PC{\bmc_t} }} \frac{\partial}{\partial C_{i,t}} \! \textit{SINR}_u\PC{\bmc_t}\!,
\end{align}
where 
\begin{align} \label{eq:sinr_gradient}
\raisetag{9.5ex}
        \frac{\partial}{\partial C_{i,t}} \textit{SINR}_u\PC{\bmc_t}  = \frac{\frac{\partial}{\partial C_{i,t}} \PR{\phi\PC{\bmc_t}} \Psi\PC{\bmc_t} - \phi\PC{\bmc_t} \frac{\partial}{\partial C_{i,t}} \PR{\Psi\PC{\bmc_t}}}{\PC{\Psi\PC{\bmc_t}}^2}.
\end{align}
The partial derivatives of the desired signal power in \fref{eq:des_signal_power}, the interference power in \fref{eq:interf_power}, the channel estimation power in \fref{eq:channel_est_power}, and the noise power in \fref{eq:noise_power} are given in \fref{eq:gradient_desired_signal}, \fref{eq:gradient_interference}, \fref{eq:gradient_chest}, and \fref{eq:gradient_noise}, respectively.
In all of these expressions, the matrix~$\bL_1\PR{t}$ is defined as in \fref{eq:l1_t}.

\subsection{Binarization Regularizer} \label{sec:regularizer}
\subsubsection{Objective Function} \label{sec:regularizer_cost_function}
For the regularizer $R\PC{\bC}$ that promotes binary-valued entries in~$\bC$, we adopt the method put forward in \cite{Castaneda2018}. 
Similarly to the objective functions in~\fref{eq:capacity_cost_function} and~\fref{eq:sum_rate_cost_function}, we want to maximize this expression. Therefore, we define the regularizer as follows:
\setcounter{equation}{27}
\begin{align} \label{eq:regularizer}
R\PC{\bC} = -\sum_{t=1}^T \sum_{u=1}^U \alpha \abs{C_{u,t} - 0.5}^2,
\end{align}
where $\alpha \geq 0$ is a regularization parameter, with larger values enforcing binary-valued solutions more strictly (see \fref{tbl:alpha_mmwave_and_cell_free}).

\subsubsection{Gradient} \label{sec:regularizer_gradient}
The partial derivatives of the regularizer in~\fref{eq:regularizer} with respect to one entry of the matrix $\bC$ are given by 
\begin{align}
    \frac{\partial R(\bC) }{\partial C_{i,t}} = -\sum_{t=1}^T \sum_{u=1}^U  \frac{\partial}{\partial C_{i,t}} \alpha \abs{C_{u,t} - 0.5}^2,
\end{align}
where
\begin{align}
\frac{\partial}{\partial C_{i,t}} \alpha \abs{C_{u,t} - 0.5}^2 = 
\begin{cases}
\alpha \PC{2 C_{u,t} - 1}, & \text{for}\  i = u,\\
0, & \text{for}\  i \neq u.
\end{cases}
\end{align}

\section{Orthogonal Projection onto Constraint Set}\label{sec:proximal_operator}
\subsection{Douglas-Rachford Splitting (DRS)}\label{sec:drs}
We now detail the implementation of $\mathrm{prox}_{\setX_{\setC_U\cap\setC_T}}(\cdot)$ in~\fref{eq:gradient_and_proximal}, which corresponds to the orthogonal projection in the intersection of two simplexes. 
To determine a solution, we use Douglas-Rachford splitting (DRS)~\cite{Douglas1956}. 
DRS alternatively projects onto the sets~$\setC_U$ and~$\setC_T$ until a solution is found. Concretely, we solve the following optimization problem:
\begin{align}\label{eq:proximal_operator}
\mathrm{prox}_{\setX_{\setC_U\cap\setC_T}} \PC{\bZ} = \underset{\bX}{\text{arg min} } \ \psi \PC{\bX,\bZ} + \xi\PC{\bX,\bZ},
\end{align}
with the two functions 
\begin{align}
\psi \PC{\bX,\bZ} & = \frac{\beta}{2} \frobnorm{\bX-\bZ}^2 + \setX_{\setC_U} \PC{\bX}  \label{eq:proximal_a} \text{, and} \\
\xi \PC{\bX,\bZ}& = \frac{\beta}{2} \frobnorm{\bX-\bZ}^2 + \setX_{\setC_T}  \PC{\bX}  \label{eq:proximal_b}.
\end{align}
Here, $\beta\geq0 $ is a tuning parameter that can be used to accelerate convergence. 
The projection onto the sets~$\setC_U$ and $\setC_T$ are enforced using indicator functions $\setX_{\setC_U}$ and $\setX_{\setC_T} $, as in \fref{eq:indicator_function}.

DRS iteratively carries out the following two steps \cite{Douglas1956}:
\begin{align} 
\bV^{(k+1)} = \, &\mathrm{prox}_{\psi}\PC{\!\bG^{(k)}\!}\! =\underset{\bX \in \setC_U  }{\argmin} \frobnorm{\bX\!-\!\frac{\beta \bZ + \bG^{(k)}}{\beta+1}}^2 \label{eq:proximal_c},\\
\bG^{(k+1)} = \, &\mathrm{prox}_{\xi}\PC{2\bV^{(k+1)}\!-\!\bG^{(k)}}+\bG^{(k)}-\bV^{(k+1)} \notag\\
= \, &\underset{\bX \in \setC_T }{\argmin}\frobnorm{\bX-\frac{\beta \bZ + \PC{2\bV^{(k+1)} - \bG^{(k)}}}{\beta+1}}^2   + \bG^{(k)} \nonumber\\ 
& \qquad \quad \,\, - \bV^{(k+1)} \label{eq:proximal_d}.
\end{align}
We initialize the procedure with $\bG^{(1)} = \boldsymbol{0}_{U \times T}$. After convergence or a maximum number of iterations $K_{\text{max}}$ is reached, DRS outputs the matrix that minimizes \fref{eq:proximal_operator}, $\bV^{(K_\text{max})}$.

\subsection{Orthogonal Projection onto Simplex with Inequality Constraints}

As \fref{eq:proximal_c} and \fref{eq:proximal_d} project each column of the input matrix onto a simplex constrained by inequalities, we require an orthogonal projection onto a general simplex specified by inequality constraints.
Given a vector $\bmq \in \opR^{M}$, our objective is to find a solution vector $\bmp^* \in \setC^{M}$ closest to $\bmq$ as follows: 
\begin{align} \label{eq:proj_with_inequality}
\underset{\bmp \in [0,1]^{M}}{\text{minimize}}  \frac{1}{2} \! \norm{\bmp-\bmq}_2^2 
\text{  subject to  }  \ell_{\text{min}} \leq \sum_{i=1}^{M} p_i \leq \ell_{\text{max}}.
\end{align}
In \fref{app:kkt_conditions}, we provide the KKT conditions \cite[Sec.~5.5.3]{Boyd2004} for the above problem. 
The KKT conditions are a set of necessary conditions that help identify a potential solution to a constrained optimization problem. 
We utilize these conditions to develop an algorithm that calculates the above projection.  

From the KKT conditions derived in \fref{app:kkt_conditions}, the solution $\bmp^*$ to \fref{eq:proj_with_inequality} can be written entrywise as 
\begin{align}
p_{i}^{*} = \max\chav{\min\chav{q_i+\lambda^*,1},0}, \ \text{for } i=1,\dots,M,
\end{align}
where $\lambda^* \in \opR$. Thus, solving \fref{eq:proj_with_inequality} is all about finding a value~$\lambda^*$ such that the resulting vector $\bmp^*$ satisfies the sum constraint while remaining as close as possible to $\bmq$ in Euclidean distance. 

We now outline the procedure to find $\lambda^*$:
\begin{enumerate}
    \item We sort $\bmq$ in descending order to generate $\tilde{\bmq}$.
    \item We obtain a new vector $\hat{\bmq}$ by subtracting the largest element of $\tilde{\bmq}$ from all entries, so that the first element is~$0$ and the other elements are negative.
    \item We build a vector $\boldsymbol{\chi} = \PR{-\hat{\bmq}^T, \boldsymbol{1}^T_{M}-\hat{\bmq}^T}^T$ so that $\bmp = \max\chav{\min\chav{\hat{\bmq}+\chi_i,1},0} \in [0,1]^{M}$, for $i=1,\dots,2M$. 
    \item We consider intervals between consecutive values of $\boldsymbol{\chi}$, e.g., $\PR{\chi_i,\chi_{i+1}}$. 
    We aim to identify in which of such intervals the constraints on $\bmp$ are met, i.e., $\ell_{\text{min}} \leq \sum_{i=1}^M \max \chav{\min \chav{\hat{q}_i+ \lambda, 1},0}   \leq \ell_{\text{max}}$, for $\lambda \in \PR{\chi_i,\chi_{i+1}}$. 
    \item By finding such search interval $\PR{\chi_i,\chi_{i+1}}$ and using the KKT conditions provided in \fref{app:kkt_conditions}, we compute candidate values for $\lambda^*$ and select the one that minimizes the MSE in~\fref{eq:proj_with_inequality}.
\end{enumerate}

\section{Quantization to Discrete Constraint Set} \label{sec:quantization_algorithm}
As described in \fref{sec:fbs}, the matrix $\bC^{(I_\text{max})}$ contains values in the range $[0,1]$, so they need to be quantized to binary values in the set $\{0,1\}$. 
First, we check if quantization is necessary by verifying whether all entries are already in $\{0,1\}$. If so, we simply set the final quantized matrix $\bC^{\text{quant}} = \bC^{(I_\text{max})}$. 
If not, we perform the following quantization strategy that maps the entries to $\{0,1\}$ while satisfying the constraints $\setC_U \cap \setC_T$.

\subsection{Projecting on Lower Bounds}

The first step of quantization enforces the lower-bound constraints: the minimum number of UEs that must transmit in each time slot ($U_{\text{min}}$) and the minimum number of time slots each UE must transmit in ($T_{\text{min}}$), as defined in \fref{eq:set_1} and \fref{eq:set_2}, respectively.
To this end, we start by setting $\bC = \bC^{(I_\text{max})}$ and initializing the set $\setU = {1,\dots,U}$ with all UEs and the matrix $\bC^{\text{lower}} = \boldsymbol{0}_{U \times T}$. 
For each time slot $t$, we select the indices of the $U_{\text{min}}$ UEs with the highest values in the $t$th column $\bmc_t$ of~$\bC$, and set the corresponding entries in matrix $\bC^{\text{lower}}$ to 1.
Then, we check which UEs have already been scheduled for at least $T_{\text{min}}$ slots in $\bC^{\text{lower}}$. 
These UEs are removed from $\setU$, and we proceed to the next time slot. This continues until all~$T$ time slots are processed.
After this lower-bound projection, we evaluate the objective function $F(\bC^{\text{lower}})$ to prepare for the next step: enforcing the upper-bound constraints.

\subsection{Projecting on Upper Bounds}
The second step of quantization enforces the upper-bound constraints: the maximum number of UEs allowed to transmit in each time slot ($U_{\text{max}}$) and the maximum number of time slots each UE is allowed to transmit in ($T_{\text{max}}$), as defined in~\fref{eq:set_1} and~\fref{eq:set_2}, respectively.
To this end, we start by setting $\bC = \bC^{(I_{\text{max}})}$, $\bC^{\text{upper}} = \bC^{\text{lower}}$, and initializing the set $\setU = {1,\dots,U}$ with all UEs. 
Then, we compute the number of additional UEs that can be scheduled in a time slot, given by $U_{\text{gap}} = U_{\text{max}} - U_{\text{min}}$.
Next, we identify the UEs that have not yet been scheduled in time slot $t$ and that are still in set $\setU$. 
Among these, we select the $U_{\text{gap}}$ UEs with the highest values in the $t$th column of $\bC^{\text{upper}}$, $\bmc^{\text{upper}}_t$, and add them to a candidate set $\setS$.
We define a test matrix $\bC^{\text{test}} = \bC^{\text{upper}}$. 
For each UE $u \in \setS$, we evaluate whether quantizing its corresponding entry of matrix $\bC^{\text{test}}$ to $1$ decreases the objective function $F(\bC^{\text{test}})$. If it does, we update the corresponding entry of $\bC^{\text{upper}}$ to $1$. 
If this update causes UE $u$ to reach its maximum allowed transmissions ($T_{\text{max}}$), we remove it from $\setU$.
This process is repeated for all $T$ time slots. After the upper-bound projection is complete, the final quantized matrix is given by $\bC^{\text{upper}}$.

\section{Results} \label{sec:simulation_results}
We now demonstrate the effectiveness of our UE scheduling framework for two simulated scenarios: (i) a centralized mmWave massive MU-MIMO scenario and (ii) a sub-6-GHz cell-free massive MU-MIMO scenario. 
Unless stated otherwise, all remaining parameters are common to both scenarios. 
The per-UE power control dynamic range is set to $\eta = 6$\,dB. 
For the noise power, we consider $N_0 = \textit{W} \times k_B \times T_0 \times \textit{NF} \  \text{[W]} $ \cite{Ngo2017}, where $\textit{W}=100$\,MHz is the bandwidth, $k_B = 1.381 \times 10^{-23}$\,J/K is the Boltzmann constant, $T_0 = 290$\,K is the noise temperature, and $\textit{NF} = 9$\,dB is the noise figure. 
We evaluate our framework using channel vectors generated with Remcom's Wireless InSite~\cite{Remcom}. 
To generate a channel realization, $U$ UE positions are selected uniformly and independently at random.
We perform $10^{3}$ channel realizations and $\textit{I}_{\text{MC}} = 10^{5}$ Monte Carlo trials per channel realization. 
For the FBS algorithm (cf. \fref{sec:fbs}), we consider an initial stepsize of $\tau^{(1)} = 0.01$ and a maximum number of iterations for the gradient step of $I_{\text{max}} = 100$. 
For the DRS algorithm (cf. \fref{sec:drs}), we consider a maximum number of iterations of $K_{\text{max}} = 2000$. 

\subsection{Performance Metrics} \label{sec:performance_metrics}
For the per-UE uncoded BER and per-UE MSE, we report our results as per-UE cumulative distribution functions (CDFs); 
for the per-UE HMI and per-UE achievable rate, we use complementary CDFs  (CCDFs). 
We also summarize some of our results by looking at the 90-percentile of the CDFs and CCDFs, which we define as $\text{Pr[X} \leq \text{x]} = 0.9$ and $\text{Pr[X} > \text{x]} = 0.9$, respectively. 
All of the considered performance metrics are evaluated after utilizing an LMMSE equalizer. 
We detail the per-UE uncoded BER, per-UE HMI, per-UE MSE, and per-UE achievable rate performance metrics next.

\subsubsection{Uncoded Per-UE BER}
The uncoded BER for the $u$th UE is defined as
\begin{align}
\textit{BER}_u = \frac{\sum_{i=1}^{\textit{I}_{\text{MC}}} \sum_{t=1}^{T_u} \varepsilon_{i,t}^u}{Q \textit{I}_{\text{MC}} T_u},
\end{align}
where $T_u$ is the number of time slots that the $u$th UE is active ($T_{\text{min}} \leq T_{\text{u}} \leq T_{\text{max}}$), $\varepsilon_{i,t}^u$ is the number of bit errors for the $u$th UE in trial $i$ and time slot $t$, and $Q$ is the number of bits per symbol. 
For QPSK and 16-QAM modulations, for example, we have $Q = 2$ and $Q = 4$ bits, respectively.

\subsubsection{Per-UE HMI}
The HMI for the $u$th UE is defined as
\begin{align}
    I_u = \sum_{i=1}^Q I(b^i_u,\hat{b}^i_u) &= \sum_{i=1}^Q H(b^i_u) - H(b^i_u|\hat{b}^i_u),
\end{align}
where $H(b^i_u) = -\sum_{b^i_u \in \{0,1\}} p(b^i_u)\log_2(p(b^i_u))$ is the entropy of $b^i_u$ and
\begin{align}
H(b^i_u|\hat{b}^i_u) = -\!\!\!\!\sum_{b^i_u \in \{0,1\}}\sum_{\hat{b}^i_u \in \{0,1\}} \!\!\!p(b^i_u,\hat{b}^i_u)\log_2\PC{\!\frac{p(b^i_u,\hat{b}^i_u)}{p(b^i_u)}\!}
\end{align}
is the conditional entropy of $b^i_u$ given $\hat{b}^i_u$. In the expressions above, $Q$ is the number of bits per symbol, $b^i_u$ is the $i$th bit transmitted by the $u$th UE, $\hat{b}^i_u$ is the $i$th estimated bit transmitted by the $u$th UE, $p(b^i_u = 0) = p(b^i_u = 1) = 1/2$, and $p(b^i_u,\hat{b}^i_u)$ is the joint probability distribution of $b^i_u$ and $\hat{b}^i_u$, which we calculate empirically. 
For the empirical calculation of the joint probability $p(b^i_u,\hat{b}^i_u)$, we only take into account when UE $u$ is active. 

\subsubsection{Per-UE MSE}
The per-UE MSE is defined as 
\begin{align}
\textit{MSE}_u = \frac{\sum_{i=1}^{I_{\text{MC}}} \sum_{t=1}^{T_u} \abs{s^i_u\PR{t}-\hat{s}^i_u\PR{t}}^2}{I_{\text{MC}} T_u},
\end{align}
where $s^i_u\PR{t}$ is the transmit symbol of UE $u$ in trial $i$ and time slot $t$, and $\hat{s}^i_u\PR{t}$ is the estimated transmit symbol of UE $u$ in trial $i$ and time slot $t$. 
Recall that $\hat{\bms}^{i}\PR{t} = \bW\PR{t}^{\text{H}} \bmy^{i} \PR{t}$, where $\bmy^{i} \PR{t}$ is the receive vector at trial $i$ and time slot $t$.

\subsubsection{Per-UE Achievable Rate}
The per-UE achievable rate is defined as
\begin{align}
\textit{Rate}_u = \sum_{t=1}^T \log_2\PC{1+\Omega_u\PC{\bmc_t} }\!,
\end{align}
where $\Omega_u\PC{\bmc_t}$ is given in \fref{eq:metric_sinr_u_t} and should not be confused with the one described in \fref{sec:lmmse_sum_rate}. 
In this expression, $\bme_j\PR{t}$ is the $j$th column of the noise matrix $\bE\PR{t} = \bE \bD_{\bC}\PR{t}$. 
\setcounter{equation}{42}
\begin{figure*}[!t]
\begin{align} 
\!\!\!\!\!\!\Omega_u\PC{\bmc_t} =  \frac{\abs{\bmw_u\PR{t}^{\text{H}} \!\hat\bmh_u\PR{t}}^2}{\sum_{j=1,j\neq u}^U \abs{\bmw_u\PR{t}^{\text{H}} \!\hat\bmh_{j}\PR{t}}^2 \!\!+\sum_{j=1}^U \abs{\bmw_u\PR{t}^{\text{H}} \! \bme_{j}\PR{t}}^2 \!\! +N_0 \vecnorm{\bmw_u\PR{t}}_2^2} \!\!\!\!
\label{eq:metric_sinr_u_t}
\end{align}
\hrule
\end{figure*}

\subsection{Existing Methods and Baseline Algorithms}\label{sec:baseline_algorithms}

In our simulation results, we refer to the post-LMMSE equalization MSE, channel capacity, and post-LMMSE equalization sum of achievable rates as ``MSE-LMMSE,'' ``channel capacity,'' and ``achievable rate,'' respectively. 

To benchmark our framework against existing methods, we also simulate the ``SUS'' \cite{Yoo2006}, ``CSS'' \cite{Choi2019}, and ``greedy'' \cite{Choi2019} algorithms. 
For these methods, we schedule the UEs for the first time slot and then continue scheduling the remaining unscheduled UEs in subsequent time slots until all UE requests are distributed across the $T$ available time slots.
In addition, we compare our approach with the low-complexity methods ``LoFi (K=1),'' ``LoFi (K=4),'' ``LoFi++ (K=1),'' and ``LoFi++ (K=4),'' proposed in \cite{Gallyas-Sanhueza2024}.
Finally, we also investigate two additional baseline methods: ``random,'' and ``all UEs active'' (``AUA'' for short).
The ``random'' baseline picks the subsets of UEs uniformly at random, and the ``AUA'' baseline considers that all UEs are scheduled in all time slots. 

Differently from our methods, in one of the baseline algorithms, ``random,'' and in the existing methods, ``LoFi (K=1),'' ``LoFi (K=4),'' ``LoFi++ (K=1),'' ``LoFi++ (K=4),'' ``SUS,'' ``CSS,'' and ``greedy,'' we consider that each UE is only allowed to transmit in one time slot, i.e., $T_{\text{min}}=T_{\text{max}}=1$ and $U/T$ UEs are allowed to transmit every time slot, i.e., $U_{\text{min}}=U_{\text{max}}=U/T$.
The reason for this consideration is that for each one of the techniques mentioned, we would have to test all possibilities of scheduling, considering the range between $T_{\text{min}}$ and $T_{\text{max}}$ and the range between $U_{\text{min}}$ and $U_{\text{max}}$. 
Therefore, we would create new algorithms that are not in the literature, which is not our goal.

\subsection{mmWave Massive MU-MIMO Scenario}
\subsubsection{Simulation Setup} \label{sec:mmwave_simulation_setup}
We consider a mmWave massive MU-MIMO system with a carrier frequency of $60$\,GHz.
In this scenario, we set the per-UE transmit power to $\rho_u = 100$ [mW] and the UEs transmit QPSK symbols with Gray mapping.
At the BS, we consider a ULA with half-wavelength antenna spacing. The BS and UE antennas are omnidirectional and are at a height of $15$\,m and $1.65$\,m, respectively. 
The channel vectors are generated for $30,351$ UE positions in an area of $150\,\text{m} \times 200\, \text{m}$. 

In \fref{tbl:mmwave_and_cell_free_scenario}, we list three different system sizes for the mmWave massive MU-MIMO scenario, S1, S2, and S3, where we vary the number of receive antennas $B$, i.e., the number of antennas at the BS $n_{\text{A}}$, the number of UEs $U$ to be served, and the number of available time slots $T$. 
Furthermore, considering the number of served UEs and available time slots, we define the minimum and maximum number of time slots per UE and the minimum and maximum number of UEs per time slot, $T_{\text{min}}$, $T_{\text{max}}$, $U_{\text{min}}$, and $U_{\text{max}}$, respectively. 
In the last column of the table, we define the number of random initializations of our UE scheduling algorithm (cf. \fref{sec:problem_relaxation}) for each one of the described scenarios. 
In \fref{tbl:alpha_mmwave_and_cell_free}, we list the regularization parameters $\alpha$ from \fref{eq:regularizer} for each one of the scenarios described in \fref{tbl:mmwave_and_cell_free_scenario}, S1, S2, and S3. 
We highlight that we consider the ES baselines only on the smallest system, i.e., Scenario S1, due to computational complexity. 
In this baseline, we test every possible scheduling matrix $\bC$ and select the one that minimizes the given objective function. 
We refer to the ES algorithm that minimizes the MSE as ``ES-MSE.'' 
We refer to the ES algorithms that maximize the channel capacity and sum of achievable rates as ``ES-capacity'' and ``ES-rate,'' respectively.

\begin{table}[!t]
\centering
\caption{Evaluated scenarios for a mmWave and a cell-free massive MU-MIMO systems}
\label{tbl:mmwave_and_cell_free_scenario}	
\renewcommand{\arraystretch}{1.1}
\resizebox{0.9\columnwidth}{!}{%
\begin{tabular}{@{}lcccccccccc@{}}
\toprule
Scenario& $B$ & $L$ & $n_{\text{A}}$& $U$ & $T$ & $T_{\text{min}}$ & $T_{\text{max}}$ & $U_{\text{min}}$ & $U_{\text{max}}$ & $K_{\text{init}}$ \\ 
\midrule
S1 & $16$ & $1$ & $16$ & $16$ & $2$ & $1$ & $2$ & $8$ & $10$ & $80$ \\ 
S2 & $32$ & $1$ & $32$ & $32$ & $2$ & $1$ & $2$ & $16$ & $20$ & $10$ \\ 
S3 & $32$ & $1$ & $32$ & $64$ & $4$ & $1$ & $2$ & $16$ & $20$ & $3$ \\ 
S4 & $16$ & $8$ &2& $16$ & $2$ & $1$ & $2$ & $8$ & $10$ & $80$ \\ 
S5 & $32$ & $16$ &2& $32$ & $2$ & $1$ & $2$ & $16$ & $20$ & $10$ \\ 
S6 & $32$& $16$ &2& $64$ & $4$ & $1$ & $2$ & $16$ & $20$ & $3$ \\ 
\bottomrule 
\end{tabular}%
}
\end{table}

\begin{table}[!t]
\centering
\caption{Values for $\alpha$ from \fref{eq:regularizer} for the mmWave and cell-free massive MU-MIMO systems}
\label{tbl:alpha_mmwave_and_cell_free}	
\renewcommand{\arraystretch}{1.1}
\resizebox{0.9\columnwidth}{!}{%
\begin{tabular}{@{}lccc@{}}
\toprule
Scenario & MSE-LMMSE & channel capacity & achievable rate\\ 
\midrule
S1 & $10^{-1}$ & $10$ & $1$\\
S2 & $10^{-1}$ & $5$ & $1$\\
S3 & $1$ & $1$ & $10^{-1}$\\
S4 & $10$ & $5$ & $5$\\
S5 & $10^{-5}$ & $20$ & $1$\\
S6 & $10^{-3}$ & $1$ & $10^{-2}$\\
\bottomrule 
\end{tabular}%
}
\end{table}

\subsubsection{Results} \label{sec:mmwave_results}
\begin{figure*}[h]
\centering
\setlength{\tabcolsep}{0pt} 
\renewcommand{\arraystretch}{0} 
\subfigure[per-UE uncoded BER]{\includegraphics[height=4cm,width=0.24\linewidth]{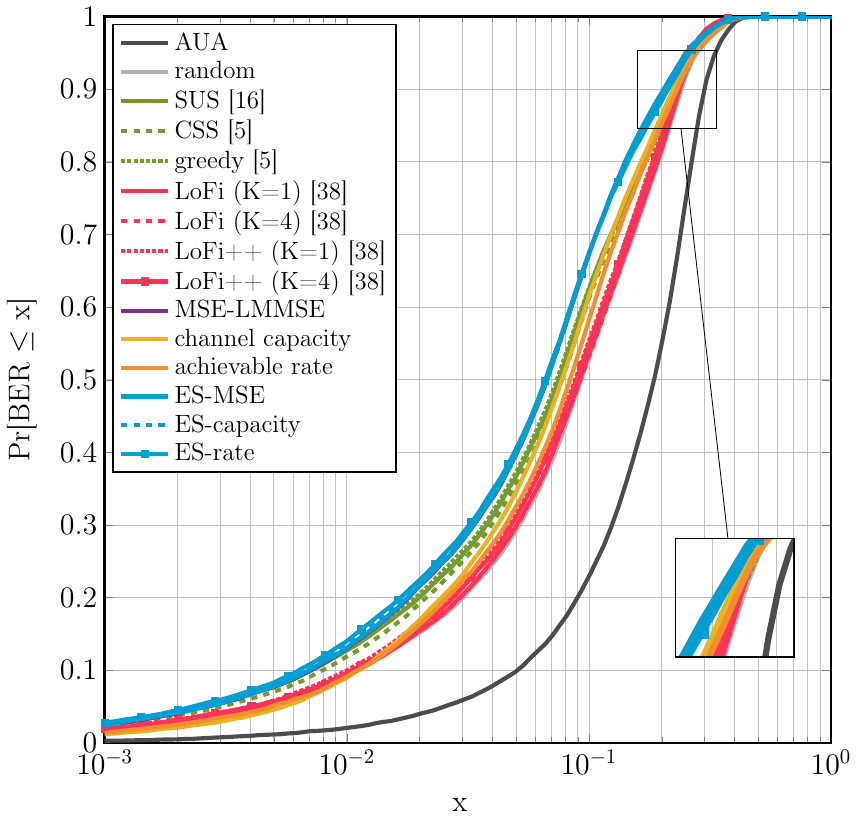}
\label{fig:ber_mmwave_16_16}}
\subfigure[per-UE HMI]{\includegraphics[height=4cm,width=0.24\linewidth]{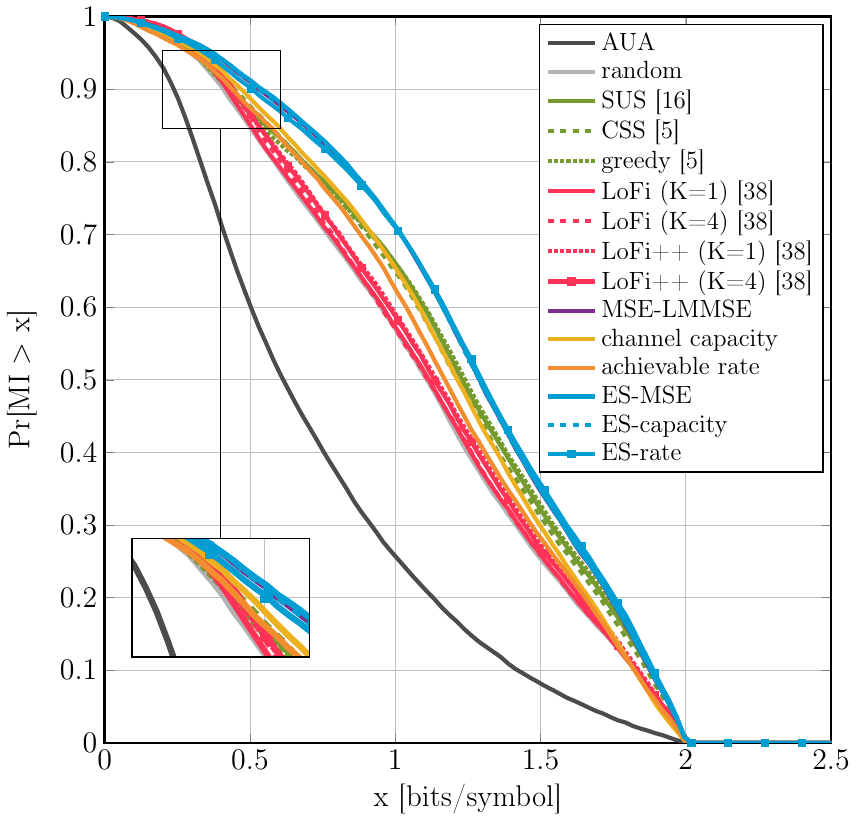}\label{fig:mi_mmwave_16_16}}
\subfigure[per-UE MSE]{\includegraphics[height=4cm,width=0.24\linewidth]{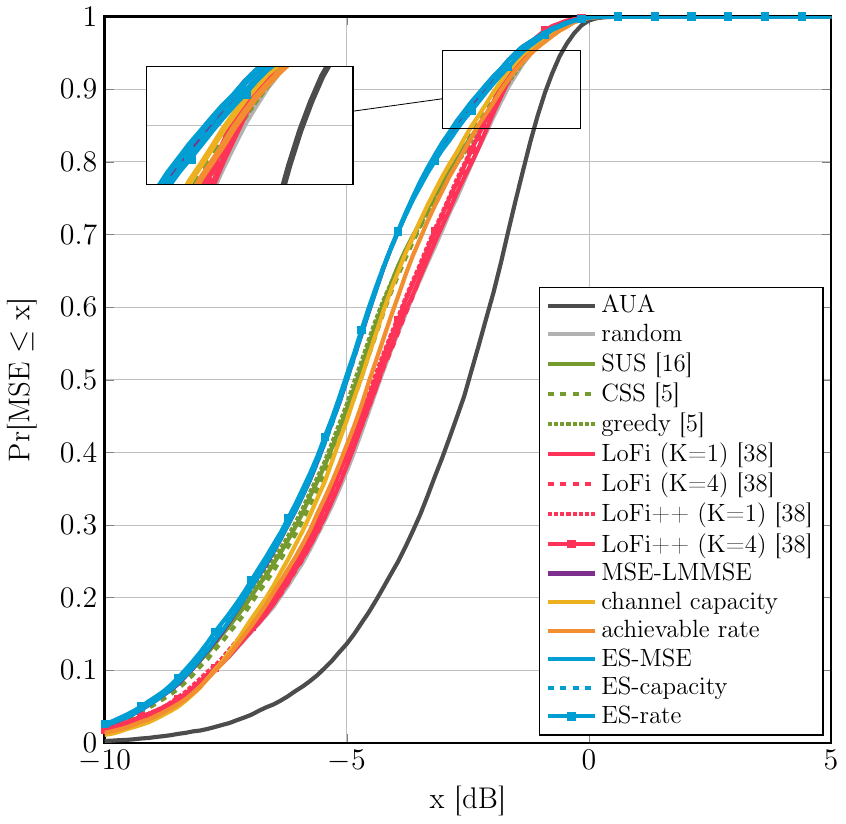}\label{fig:mse_mmwave_16_16}}
\subfigure[per-UE achievable rate]{\includegraphics[height=4cm,width=0.24\linewidth]{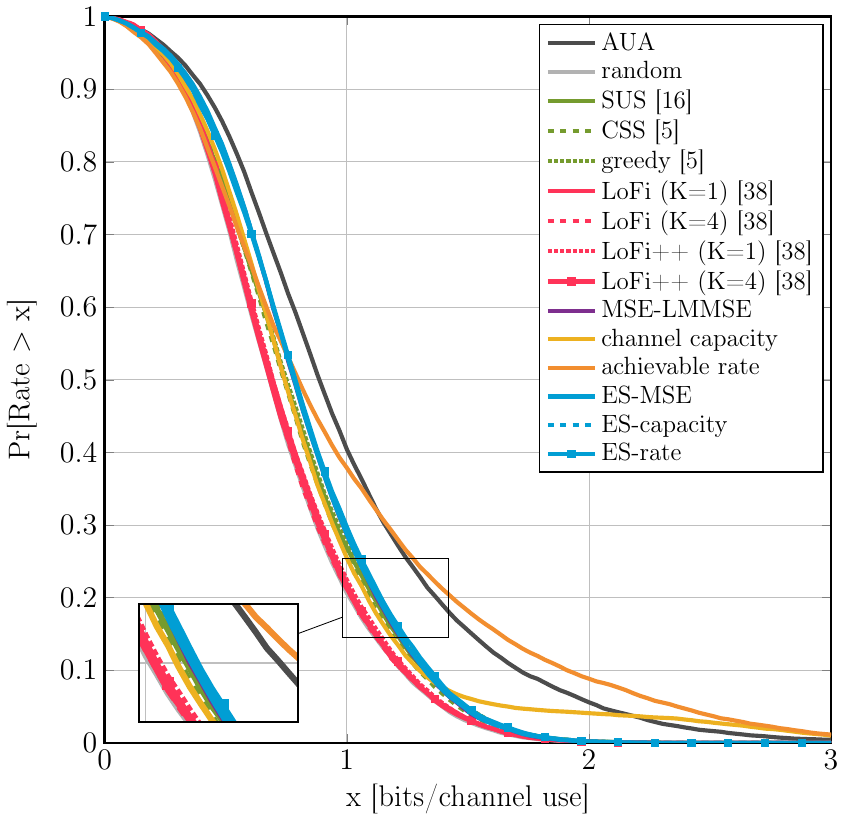}\label{fig:rate_mmwave_16_16}}
\caption{BER (a), HMI (b), MSE (c), and per-UE achievable rate (d) performance for a mmWave massive MU-MIMO system in Scenario S1: $B=16$ receive antennas, $L = 1$ AP, $n_{\text{AP}} = 16$ antennas per AP, $U=16$ UEs, $T=2$ time slots, $U_{\text{min}} = 8$ UEs, $U_{\text{max}} = 10$ UEs, $T_{\text{min}} = 1$ time slot, and $T_{\text{max}} = 2$ time slots.}
\label{fig:mmwave_16_16}
\end{figure*}

In \fref{fig:mmwave_16_16}, we show the results for Scenario S1, whose parameters are specified in \fref{tbl:mmwave_and_cell_free_scenario}. 
In Figs.~\ref{fig:ber_mmwave_16_16}-(c), we show that  ``MSE-LMMSE'' has the best performance for $90\%$ of the UEs, together with ``ES-MSE,'' ``ES-capacity,'' and ``ES-rate.'' 
As expected, the ``AUA'' baseline shows the worst performance in terms of per-UE BER, HMI, and MSE metrics, as shown in Figs.~\ref{fig:ber_mmwave_16_16}-(c), while achieving the best result in terms of per-UE achievable rate, as shown in \fref{fig:rate_mmwave_16_16}.
We highlight that none of our methods, ``MSE-LMMSE,'' ``channel capacity,'' and ``achievable rate'' optimize the presented metrics, per-UE BER, HMI, MSE, and achievable rate. 
In the case of ``MSE-LMMSE,'' for example, we are optimizing the sum of MSEs over all UEs and time slots, but we are showing results for the per-UE MSE. 
Therefore, we cannot guarantee that ``MSE-LMMSE'' has the best performance for the per-UE MSE and that ``achievable rate'' has the best performance for the per-UE achievable rate. We discuss this issue in \fref{sec:limitations}.

In \fref{tbl:mmwave_results}, we summarize the results for Scenario S2 (left) and Scenario S3 (right). 
The parameters of these scenarios are specified in \fref{tbl:mmwave_and_cell_free_scenario}. 
In both cases, we focus on the 90-percentile (cf. \fref{sec:performance_metrics}) of our performance metrics, per-UE BER, HMI, MSE, and achievable rate.

In \fref{tbl:mmwave_results} (left), we see that the method with the best overall performance, ``MSE-LMMSE,'' enables for $90\%$ of the UEs, a BER of less than $20.3\%$, an HMI of at least $0.546$ bits/symbol, an MSE $0.319$ dB lower than the ones from the best existing methods seen in this metric, ``CSS'' and ``LoFi++ (K=1),'' and a per-UE achievable rate of at least $0.379$ bits/channel use. 
As expected, the ``AUA'' baseline shows the worst performance in terms of per-UE BER, HMI, and MSE metrics, while achieving the best results in terms of per-UE achievable rate.

In \fref{tbl:mmwave_results} (right), we show that the method with the best BER, HMI, and MSE performance, ``achievable rate,''  enables for $90\%$ of the UEs, a BER of less than $19.1\%$, an HMI of at least $0.6$ bits/symbol, and an MSE $0.284$ dB lower than the one from the best existing method seen in this metric, ``CSS,'' and a per-UE achievable rate of at least $0.206$ bits/channel use.
As expected, the ``AUA'' baseline shows the worst performance in terms of per-UE BER, HMI, and MSE metrics, while achieving the best results in terms of per-UE achievable rate.
The high per-UE achievable rate performance of ``AUA'' was also observed in \cite{Palhares2022}, where it demonstrated a good average per-UE rate performance in the low-signal-to-noise ratio (SNR) regime.

\begin{table*}
\centering
\caption{90th percentile performance for two mmWave massive MU-MIMO systems. Both systems have $B=32$ receive antennas, $L = 1$ AP, $n_{\text{AP}} = 32$ antennas per AP, $U_{\textnormal{min}} = 16$ UEs, $U_{\textnormal{max}} = 20$ UEs, $T_{\textnormal{min}} = 1$ time slots, and $T_{\textnormal{max}} = 2$ time slots.}
\label{tbl:mmwave_results}
\resizebox{0.93\textwidth}{!}{
\begin{tabular}{@{}lcccc|cccc@{}}
\toprule
\multirow{2}{*}{} & \multicolumn{4}{c|}{Scenario S2: $U=32$ UEs, $T=2$ time slots} & \multicolumn{4}{c}{Scenario S3: $U=64$ UEs, $T=4$ time slots} \\
Method & BER & HMI [bits/symbol] & MSE [dB] & Rate [bits/ch] & BER & HMI [bits/symbol] & MSE [dB] & Rate [bits/ch] \\
\midrule
AUA & 0.292 & 0.260 & -1.026 & \textbf{0.432} & 0.334 & 0.163 & -0.675 & \textbf{0.281} \\
random & 0.232 & 0.438 & -1.835 & 0.333 & 0.219 & 0.484 & -2.075 & 0.184 \\
SUS \cite{Yoo2006} & 0.230 & 0.444 & -1.840 & 0.329 & 0.213 & 0.505 & -2.130 & 0.181 \\
CSS \cite{Choi2019} & 0.225 & 0.462 & -1.922 & 0.338 & 0.208 & 0.525 & -2.205 & 0.186 \\
greedy \cite{Choi2019} & 0.233 & 0.432 & -1.794 & 0.326 & 0.219 & 0.483 & -2.042 & 0.177 \\
LoFi (K=1) \cite{Gallyas-Sanhueza2024} & 0.231 & 0.440 & -1.843 & 0.336 & 0.219 & 0.486 & -2.083 & 0.183 \\
LoFi (K=4) \cite{Gallyas-Sanhueza2024} & 0.230 & 0.444 & -1.864 & 0.335 & 0.218 & 0.487 & -2.090 & 0.183 \\
LoFi++ (K=1) \cite{Gallyas-Sanhueza2024} & 0.226 & 0.459 & -1.922 & 0.344 & -- & -- & -- & -- \\
LoFi++ (K=4) \cite{Gallyas-Sanhueza2024} & 0.227 & 0.452 & -1.896 & 0.342 & -- & -- & -- & -- \\
MSE-LMMSE & \textbf{0.203} & \textbf{0.546} & \textbf{-2.241} & 0.379 & 0.198 & 0.580 & -2.430 & 0.206 \\
channel capacity & 0.234 & 0.431 & -1.791 & 0.319 & \textbf{0.191} & 0.593 & -2.485 & 0.209 \\
achievable rate & 0.227 & 0.455 & -1.888 & 0.330 & \textbf{0.191} & \textbf{0.600} & \textbf{-2.489} & 0.206 \\
\bottomrule
\end{tabular}}
\end{table*}

\subsection{Cell-Free Massive MU-MIMO Scenario}
\subsubsection{Simulation Setup}
We consider a cell-free massive MU-MIMO system operating at a carrier frequency of $1.9$\,GHz.
In this system, we set the per-UE transmit power to $\rho_u = 10$ [mW] and the UEs transmit 64-QAM symbols with Gray mapping.
At the APs, we consider ULAs with half-wavelength antenna spacing. The APs and UE antennas are omnidirectional and are at a height of $15$\,m and $1.65$\,m, respectively. 
The channel vectors are generated for $26,934$ UE positions in an area of $200\,\text{m} \times 300\, \text{m}$. 

In \fref{tbl:mmwave_and_cell_free_scenario}, we list three different system sizes for the cell-free massive MU-MIMO scenario, S4, S5, and S6, where we vary the number of APs $L$, the number of antennas per AP $n_{\text{A}}$, the number of UEs $U$ to be served, and the number of available time slots $T$. 
Furthermore, considering the number of served UEs and available time slots, we define the minimum and maximum number of time slots per UE and the minimum and maximum number of UEs per time slot, $T_{\text{min}}$, $T_{\text{max}}$, $U_{\text{min}}$, and $U_{\text{max}}$, respectively. 
In the last column of the table, we define the number of random initializations of our UE scheduling algorithm (cf. \fref{sec:problem_relaxation}) for each one of the described scenarios.
In \fref{tbl:alpha_mmwave_and_cell_free}, we list the regularization parameters $\alpha$ from \fref{eq:regularizer} for each one of the scenarios described in \fref{tbl:mmwave_and_cell_free_scenario}, S4, S5, and S6. 
We highlight that we consider the ES baselines only on the smallest system, i.e., Scenario S4, due to computational complexity. 
In this baseline, we test every possible scheduling matrix $\bC$ and select the one that minimizes the given objective function. 
We use the same naming scheme as the one presented in \fref{sec:mmwave_simulation_setup} to refer to the ES algorithms in the cell-free massive MU-MIMO scenarios.

\subsubsection{Results}
\begin{figure*}[h]
\centering
\setlength{\tabcolsep}{0pt} 
\renewcommand{\arraystretch}{0} 
\subfigure[per-UE uncoded BER]{\includegraphics[height=4cm,width=0.24\linewidth]{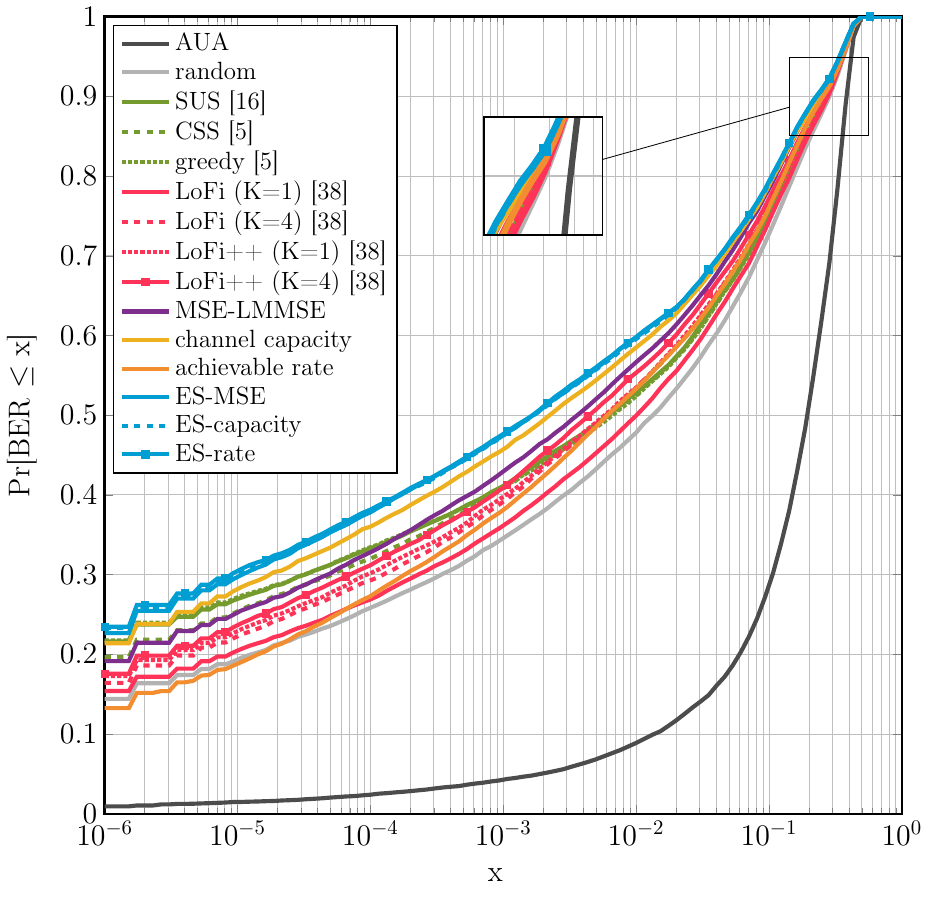}\label{fig:ber_cell_free_16_16}}
\subfigure[per-UE HMI]{\includegraphics[height=4cm,width=0.24\linewidth]{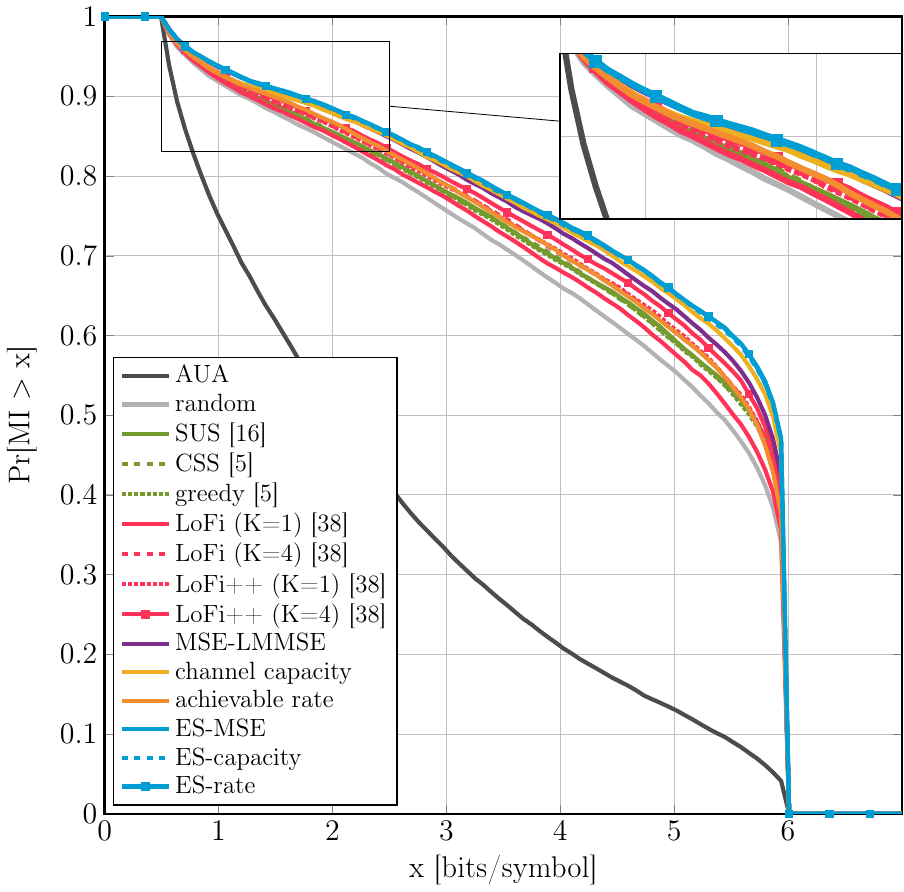}\label{fig:mi_cell_free_16_16}}
\subfigure[per-UE MSE]{\includegraphics[height=4cm,width=0.24\linewidth]{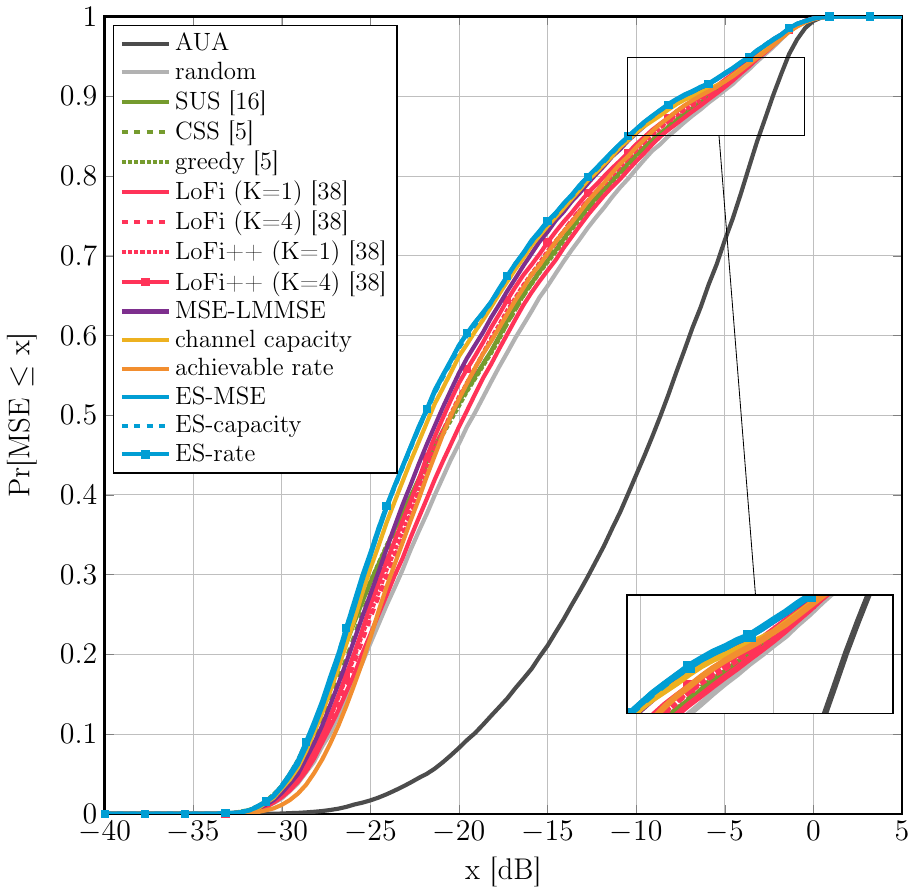}\label{fig:mse_cell_free_16_16}}
\subfigure[per-UE achievable rate]{\includegraphics[height=4cm,width=0.24\linewidth]{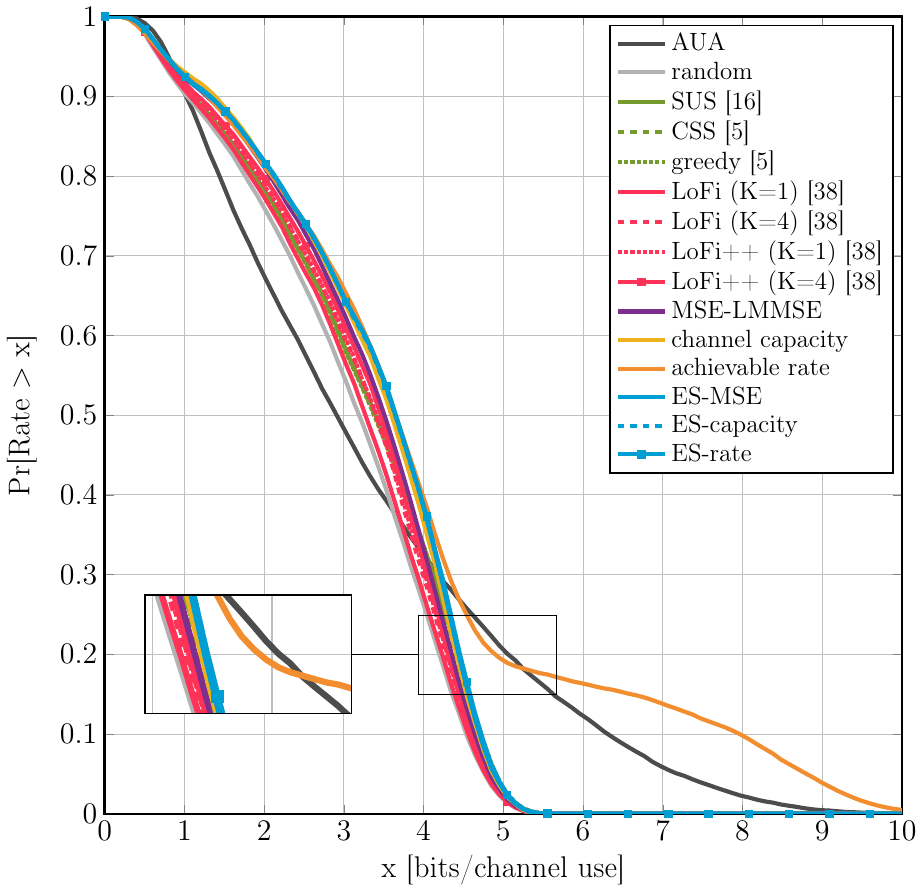}\label{fig:rate_cell_free_16_16}}
\caption{BER (a), HMI (b), MSE (c), and per-UE achievable rate (d) performance for a cell-free MU-MIMO system in Scenario S4: $B=16$ receive antennas, $L = 8$ APs, $n_{\text{AP}} = 2$ antennas per AP, $U=16$ UEs, $T=2$ time slots, $U_{\text{min}} = 8$ UEs, $U_{\text{max}} = 10$ UEs, $T_{\text{min}} = 1$ time slot, and $T_{\text{max}} = 2$ time slots.}
\label{fig:cell_free_16_16}
\end{figure*}

In \fref{fig:cell_free_16_16}, we show the results for Scenario S4, whose parameters are specified in \fref{tbl:mmwave_and_cell_free_scenario}. In Figs.~\ref{fig:ber_cell_free_16_16}-(d), we show that  ``MSE-LMMSE'' has the best performance for $90\%$ of the UEs, together with ``ES-MSE'', ``ES-Capacity,'' and ``ES-rate.''
As expected, the ``AUA'' baseline shows the worst performance in terms of per-UE BER, HMI, MSE, and achievable rate metrics, as shown in Figs.~\ref{fig:ber_cell_free_16_16}-(d).
We highlight that none of our methods, ``MSE-LMMSE,'' ``channel capacity,'' and ``achievable rate'' optimize the presented metrics, per-UE BER, HMI, MSE, and achievable rate, as explained in \fref{sec:mmwave_results}. 

In \fref{tbl:cell_free_results}, we summarize the results for Scenario S5 (left) and Scenario S6 (right), respectively. 
The parameters of these scenarios are specified in \fref{tbl:mmwave_and_cell_free_scenario}. 
In both cases, we look at the 90-percentile (cf. \fref{sec:performance_metrics}) of our performance metrics, per-UE BER, HMI, MSE, and achievable rate. 

In \fref{tbl:cell_free_results} (left), we show that the method with the best overall performance, ``MSE-LMMSE,'' enables for $90\%$ of the UEs a BER of less than $0.2\%$, an HMI of at least $5.873$ bits/symbol, an MSE $0.642$ dB lower than the one from the best existing method seen in this metric, ``LoFi++ (K=4),'' and a per-UE achievable rate of at least $3.625$ bits/channel use. 
As expected, the ``AUA'' baseline shows the worst performance in terms of per-UE BER, HMI, MSE, and achievable rate metrics.

In \fref{tbl:cell_free_results} (right), we show that the methods with the best overall performance, ``MSE-LMMSE'' and ``achievable rate,''  enable for $90\%$ of the UEs, a BER of less than $0.6\%$ and $0.5\%$, an HMI of at least $5.759$ and $5.801$ bits/symbol, an MSE $0.919$ and $0.821$ dB lower than the one from the best existing method seen in this metric, ``CSS,'' and a per-UE achievable rate of at least $1.739$ and $1.739$ bits/channel use, respectively. 
As expected, the ``AUA'' baseline shows the worst performance in terms of per-UE BER, HMI, MSE, and achievable rate metrics.
The weak per-UE achievable rate performance of ``AUA'' was also observed in \cite{Palhares2022}, where it demonstrated a weak average per-UE rate performance in the high-SNR regime.

\subsection{Summary}
From these results, we draw the following conclusions. Even in overloaded scenarios (S3 and S6), our UE scheduling method achieves a performance comparable to S2 and S5, respectively. 
Although S3 and S6 have more UEs than receive antennas, the larger number of available time slots maintains good QoS. 
This indicates that the ratio between $B$ and $U$ is less critical than the ratio between $B$ and the number of active UEs, which in S2, S3, S5, and S6 is constrained between $U_\text{min}=16$ and $U_\text{max}=20$ for $B=32$ receive antennas.

The relatively poor performance in the mmWave massive MU-MIMO scenarios is caused by unfavorable channel conditions at these frequencies, leading to low-SNR operation; incorporating UE-side beamforming would mitigate this, but at the cost of additional complexity---this is left for future work. 
In contrast, the cell-free massive MU-MIMO scenarios perform significantly better due to the fact that UEs are likely to be in close proximity to at least one AP, enabling operation in a higher-SNR regime than centralized BS deployments.

The optimization objective strongly influences the obtained UE schedules. For example, the ``MSE-LMMSE'' objective favors scheduling the minimum number of UEs per time slot, $U_\text{min}$, to reduce estimation errors, whereas ``channel capacity'' benefits from more simultaneous transmissions, pushing toward $U_\text{max}$. 
The ``achievable rate'' reflects a trade-off between assigning more time slots per UE and limiting inter-UE interference within each slot.

Compared to the low-complexity methods ``LoFi'' and ``LoFi++'' \cite{Gallyas-Sanhueza2024}, our method achieves clear performance gains in both mmWave and cell-free massive MU-MIMO systems. 
Moreover, ``LoFi'' and ``LoFi++'' were presented in \cite{Gallyas-Sanhueza2024} for a single objective function and lack flexibility in resource allocation, whereas our framework supports multiple objectives and flexible constraints on UEs per time slot and time slots per UE. 
Their objective functions also ignore channel estimation errors. Although ``LoFi++'' improves upon ``LoFi'', it introduces additional limitations, including reduced scalability beyond two time slots ($T>2$) and no support for parallelization.

\begin{table*}
\centering
\caption{90th percentile performance for two cell-free massive MU-MIMO systems. Both systems have $B=32$ receive antennas, $L = 16$ APs, $n_{\text{AP}} = 2$ antennas per AP, $U_{\text{min}} = 16$ UEs, $U_{\text{max}} = 20$ UEs, $T_{\text{min}} = 1$ time slots, and $T_{\text{max}} = 2$ time slots.}
\label{tbl:cell_free_results}
\resizebox{0.93\textwidth}{!}{
\begin{tabular}{@{}lcccc|cccc@{}}
\toprule
\multirow{2}{*}{} & \multicolumn{4}{c|}{Scenario S5: $U=32$ UEs, $T=2$ time slots} & \multicolumn{4}{c}{Scenario S6: $U=64$ UEs, $T=4$ time slots} \\
Method & BER & HMI [bits/symbol] & MSE [dB] & Rate [bits/ch] & BER & HMI [bits/symbol] & MSE [dB] & Rate [bits/ch] \\
\midrule
AUA & $0.275$ & $1.257$ & $-5.742$ & $2.03$ & $0.421$ & $0.517$ & $-1.647$ & $0.555$ \\
random & $0.013$ & $5.394$ & $-19.209$ & $3.199$ & $0.021$ & $5.131$ & $-18.314$ & $1.525$ \\
SUS \cite{Yoo2006} & $0.009$ & $5.581$ & $-19.96$ & $3.321$ & $0.008$ & $5.592$ & $-20.033$ & $1.663$ \\
CSS \cite{Choi2019} & $0.008$ & $5.602$ & $-20.044$ & $3.339$ & $0.008$ & $5.593$ & $-20.034$ & $1.664$ \\
greedy \cite{Choi2019} & $0.009$ & $5.558$ & $-19.863$ & $3.304$ & $0.009$ & $5.557$ & $-19.883$ & $1.652$ \\
LoFi (K=1) \cite{Gallyas-Sanhueza2024} & $0.009$ & $5.555$ & $-19.85$ & $3.301$ & $0.016$ & $5.315$ & $-18.932$ & $1.575$ \\
LoFi (K=4) \cite{Gallyas-Sanhueza2024} & $0.005$ & $5.735$ & $-20.747$ & $3.45$ & $0.011$ & $5.502$ & $-19.64$ & $1.633$ \\
LoFi++ (K=1) \cite{Gallyas-Sanhueza2024} & $0.005$ & $5.731$ & $-20.74$ & $3.448$ & -- & -- & -- & -- \\
LoFi++ (K=4) \cite{Gallyas-Sanhueza2024} & $0.004$ & $5.8$ & $-21.171$ & $3.522$ & -- & -- & -- & -- \\
MSE-LMMSE & $\bf 0.002$ & $\bf 5.873$ & $\bf -21.813$ & $\bf 3.625$ & $0.006$ & $5.759$ & $\bf -20.953$ & $\bf 1.739$ \\
channel capacity & $0.005$ & $5.745$ & $-20.821$ & $3.461$ & $0.015$ & $5.349$ & $-19.061$ & $1.586$ \\
achievable rate & $0.003$ & $5.82$ & $-21.349$ & $3.585$ & $\bf 0.005$ & $\bf 5.801$ & $-20.855$ & $\bf 1.739$ \\
\bottomrule
\end{tabular}
}
\end{table*}

\section{Limitations and Future Work} \label{sec:limitations}
We summarize the limitations of the proposed UE scheduling framework. 
Although our optimization-based scheduling approach offers clear advantages over existing methods, such as flexibility in supported constraints and objective functions, and a global solution that jointly assigns UEs to time slots, it entails obvious drawbacks. 
First, there is a trade-off between flexibility and computational efficiency: compared to greedy or heuristic methods, our framework can incur higher complexity; see, e.g.,~\cite[Tbl.~I]{Gallyas-Sanhueza2024} for a comparison. 
Since our UE scheduling framework outperforms all of the considered baselines (except ES) and existing methods at the cost of increased complexity, our method may serve to generate high-quality training data for ML-based scheduling approaches \cite{Chukhno2021,Zou2021,Hsu2024,Feres2023}. 
Furthermore, since our framework minimizes a nonconvex objective function, gradient-descent-based methods may converge only to a local minimum, and no optimality guarantees can be provided. 
Thus, to assess the performance of our methods, we compared them with an ES approach.

Another point is that the objective functions considered here optimize metrics across all UEs and time slots without explicitly addressing fairness. 
More practical designs may prioritize UEs with poor channel conditions, e.g., via proportional fairness, which balances instantaneous achievable rate and long-term average throughput \cite{Kushner2004,Zhang2022,Kelly1998}. 
Integrating such objective functions into our framework is left for future work.

Finally, while UE-centric architectures are common in cell-free massive MU-MIMO systems, our framework was not developed for this setting, though an extension appears feasible. 
Frequency scheduling, addressed in prior work, is another natural extension. 
In addition, our approach does not incorporate a joint power control scheme that accounts for UE scheduling; given the difficulty of jointly optimizing power control and scheduling, this is also left for future research.

\section{Conclusions} \label{sec:conclusions}
We have proposed a novel optimization-based UE scheduling framework for MU-MIMO systems. Our framework supports a range of objective functions and constraints that specify the UE resource allocation. 
We have detailed all aspects of our problem formulation, including the presentation of the objective functions and their respective gradients, and the derivation of the proximal operator.
We have evaluated the performance of our framework using realistic mmWave and sub-6-GHz cell-free massive MU-MIMO channel vectors. 

In a small MU-MIMO mmWave scenario with $B=16$ and $U=16$, we have observed that the performance of the proposed methods often approach that of an exhaustive search and outperform all of the considered baseline algorithms as well as the existing methods in terms of per-UE BER, HMI, and MSE.
In larger systems with $U = B$, we have shown that the proposed methods outperform all of the considered baseline algorithms and the existing methods in terms of per-UE BER, HMI, and MSE.
In an overloaded scenario where $U > B$, we have demonstrated that the performance can be similar to the one with $U = B$ if the UEs' requests are spread over more time slots. 

In a small cell-free scenario with $B=16$ and $U=16$, we have observed that the performance of the proposed methods are the closest ones from an exhaustive search and outperform both the baseline algorithms as well as the existing methods in terms of per-UE BER, HMI, MSE, and achievable rate.
In larger systems with $U = B$, we have shown that the proposed methods outperform the baseline algorithms and the existing methods in terms of per-UE BER, HMI, MSE, and achievable rate.
In an overloaded scenario where $U > B$, we have demonstrated that the performance can be similar to the one with $U = B$ if the UEs' requests are spread over more time slots. 
Moreover, as seen in the cell-free massive MU-MIMO scenarios with UE scheduling, we have observed an improvement when spreading not only the UEs over more time slots but also the APs over a large area, due to the UE-AP proximity.
\vspace{-0.3cm}
\appendices
\section{Karush-Kuhn-Tucker (KKT) Conditions of (36)} \label{app:kkt_conditions}
The KKT conditions of the problem presented in \fref{eq:proj_with_inequality} are:
\begin{enumerate}
\item $p_i - q_i - \mu_i+ \upsilon_i - \gamma + \Gamma = 0, i=1,\dots,\theta$,
\item $\mu_i p_i = 0, \upsilon_i\PC{p_i-1} = 0, \gamma \PC{\ell_{\text{min}} - \sum_{i=1}^{M} p_i} = 0, \Gamma \PC{\sum_{i=1}^{M} p_i - \ell_{\text{max}}} = 0, i=1,\dots,M$,
\item $p_i \geq 0, p_i \leq 1, \sum_{i=1}^{M} p_i \geq \ell_{\text{min}}, \sum_{i=1}^{M} p_i \leq \ell_{\text{max}}, i=1,\dots,M$,\label{eq:condition_3}
\item $\mu_i \geq 0, \upsilon_i \geq 0, \gamma \geq 0, \Gamma \geq 0, i=1,\dots,M$.
\end{enumerate}
 
The variable $p_i$ for $i=1,\dots,M$, can fall in one of three cases: $p_i = 1$, $p_i \neq 0,1$, and $p_i = 0$. For $p_i \neq 0,1$, we get that $\mu_i=0$ and $\upsilon_i=0$. 
Therefore, from the third condition we have that:
    \begin{enumerate}
        \item $\ell_{\text{min}}-p_i < \sum_{j=1,j \neq i}^M p_j < \ell_{\text{max}}-p_i$ leading to $\gamma = 0$, $\Gamma = 0$, and $p_i=q_i$,
        \item $\sum_{j=1,j \neq i}^M p_j = \ell_{\text{min}}-p_i$ leading to $\Gamma= 0$ and $p_i=q_i+\gamma=0$, 
        \item $\sum_{j=1,j \neq i}^M p_j = \ell_{\text{max}}-p_i$ leading to $\gamma = 0$ and $p_i=q_i-\Gamma=0$. 
    \end{enumerate}
 
According to \cite{Duchi2008}, for $q_i > q_j$, if $p_i=0$, then $p_j=0$. 
Similarly, in this case, for $q_i < q_j$, if $p_i=1$, then $p_j=1$. 
By sorting $\bmq$ in descending order, we have $\mathbf{q} = \PR{q_1,q_2,\dots,q_M}, \text{ where } q_1 > \dots > q_M$.
If we sort $\bmp$ according to $\bmq$, we obtain
\begin{align}
     \mathbf{p} &= \PR{\underbrace{1,\dots,1}_{M_1},\underbrace{p_{M_1+1},\dots,p_{M-M_0}}_{M_1+1,\dots,M-M_0},\underbrace{0,\dots,0}_{M_0}},
\end{align}
where $M_1$ and $M_0$ are the number of elements in $\bmp$ equal to 1 and 0, respectively.

Reformulating the problem when $p_i \neq \{0,1\}$ and applying on the inequality constraints, we obtain the solution for the problem as follows:
\begin{enumerate}
\item $\ell_{\text{min}} < p_i + \sum_{j=1,j\neq i}^{M} p_j < \ell_{\text{max}}$, which leads to $p_i=q_i$,
\item $p_i + \sum_{j=1,j\neq i}^{M} p_j = \ell_{\text{min}}$ which leads to $p_i = q_i + \frac{1}{M-M_0-M_1}\PC{\ell_{\text{min}} - M_1 - \sum_{j=M_1+1}^{M-M_0} q_j}$,
\item $p_i + \sum_{j=1,j\neq i}^{M} p_j = \ell_{\text{max}}$ which leads to $p_i = q_i + \frac{1}{M-M_0-M_1}\PC{\ell_{\text{max}} - M_1- \sum_{j=M_1+1}^{M-M_0} q_j}$,
\end{enumerate}
where $i=M_1+1,\dots,M-M_0$.

\balance
\bibliographystyle{IEEEtran}
\bibliography{bib/VIPabbrv,bib/confs-jrnls,bib/publishers,bib/26TWC_UE_Scheduling}

\balance

\end{document}